%% file: main.tex
\documentclass[11pt, a4paper, superscriptaddress]{revtex4-1}
\usepackage[T2A]{fontenc}
\usepackage[english]{babel}
\usepackage[utf8]{inputenc}

\usepackage{amssymb, amsmath, enumerate, graphicx, subfigure, geometry, titlesec, setspace}

\DeclareMathOperator{\erfc}{erfc}

\usepackage[linktocpage,unicode]{hyperref}

\usepackage{subfigure,xcolor}     

\renewcommand\thesection{\arabic{section}}
\renewcommand\thesubsection{\arabic{subsection}}
\renewcommand\thesubsubsection{\arabic{subsubsection}}

\titleformat{\section}[hang]{\bfseries\Large\centering}{\thesection.}{7pt}{\Large}
\titleformat{\subsection}[hang]{\bfseries\large\filright}{\thesection.\thesubsection}{7pt}{\large}

\def\({\left(}
\def\){\right)}

\definecolor{viol}{RGB}{140,0,140}

\begin{document}

\title{Clusters of Primordial Black Holes}

\author{Konstantin~M.~Belotsky}
\affiliation{National Research Nuclear University MEPhI (Moscow Engineering Physics Institute),\\ 115409, Kashirskoe shosse 31, Moscow, Russia}
\author{Vyacheslav~I.~Dokuchaev}
\affiliation{National Research Nuclear University MEPhI (Moscow Engineering Physics Institute),\\ 115409, Kashirskoe shosse 31, Moscow, Russia}
\affiliation{Institute for Nuclear Research, Russian Academy of Sciences,\\ 117312, pr. 60-letiya Oktyabrya 7a, Moscow, Russia}
\author{Yury~N.~Eroshenko}
\affiliation{Institute for Nuclear Research, Russian Academy of Sciences,\\ 117312, pr. 60-letiya Oktyabrya 7a, Moscow, Russia}
\author{Ekaterina~A.~Esipova}
\affiliation{National Research Nuclear University MEPhI (Moscow Engineering Physics Institute),\\ 115409, Kashirskoe shosse 31, Moscow, Russia}
\author{Maxim~Yu.~Khlopov}
\affiliation{Institute of Physics, Southern Federal University\\ 344090, Stachki 194, Rostov on Don, Russia}
\author{Leonid~A.~Khromykh}
\affiliation{National Research Nuclear University MEPhI (Moscow Engineering Physics Institute),\\ 115409, Kashirskoe shosse 31, Moscow, Russia}
\author{Alexander~A.~Kirillov}
\affiliation{National Research Nuclear University MEPhI (Moscow Engineering Physics Institute),\\ 115409, Kashirskoe shosse 31, Moscow, Russia}
\author{Valeriy~V.~Nikulin}
\affiliation{National Research Nuclear University MEPhI (Moscow Engineering Physics Institute),\\ 115409, Kashirskoe shosse 31, Moscow, Russia}
\author{Sergey~G.~Rubin}
\affiliation{National Research Nuclear University MEPhI (Moscow Engineering Physics Institute),\\ 115409, Kashirskoe shosse 31, Moscow, Russia}
\affiliation{N.~I.~Lobachevsky Institute of Mathematics and Mechanics, Kazan  Federal  University, \\ 420008, Kremlevskaya  street  18,  Kazan,  Russia}
\author{Igor~V.~Svadkovsky}
\affiliation{National Research Nuclear University MEPhI (Moscow Engineering Physics Institute),\\ 115409, Kashirskoe shosse 31, Moscow, Russia}

\begin{abstract}
The Primordial Black Holes (PBHs) are a well-established probe for new physics in the very early Universe. We discuss here the possibility of PBH agglomeration into clusters that may have several prominent observable features. The clusters can form due to closed domain walls appearance in the natural and hybrid inflation models whose subsequent evolution leads to PBH formation. The dynamical evolution of such clusters discussed here is of crucial importance. Such a model inherits all the advantages of uniformly distributed  PBHs, like possible explanation of supermassive black holes existence (origin of the early quasars), the binary black hole mergers registered by LIGO/Virgo through gravitational waves, which could provide ways to test the model in future, the contribution to reionization of the Universe. If PBHs form clusters, they could alleviate or completely avoid existing constraints on the abundance of uniformly distributed PBHs, thus allowing PBH to be a viable dark matter candidate. Most of the existing constraints on uniform PBH density should be re-considered to the case of PBH clustering. Furthermore, unidentified cosmic gamma-ray point-like sources could be (partially) accounted. We conclude that models leading to PBH clustering are favored compared to models predicting the uniform distribution of PBHs.
\end{abstract}


\maketitle

\tableofcontents

\section{Introduction}
The idea of the Primordial Black Holes (PBHs) formation was predicted five decades ago \cite{p41}. Then this scenario was grounded  and developed  in \cite{p22,Carr:1974nx,CHAPLINE_pbh}
and in many other works. The PBHs have not been identified in observations, although some astrophysical effects can be attributed to the PBHs, e.g., supermassive black holes in early quasars. Therefore till now, the PBHs give information about processes in the early Universe only in the form of restrictions on the primordial perturbations \cite{p24} and on physical conditions at different epochs. There are also hopes that the evidence about PBHs existence will be found soon. It is important now to describe and develop in detail models of PBHs formation and their possible effects in cosmology and astrophysics.  

A new wave of interest to PBHs has been induced by the recent discovery of gravitational waves from BH mergers, raising discussions of PBHs with mass $10\div 100 M_{\odot}$ as dark matter (DM) candidate (see, e.g., \cite{2016PhRvL.116t1301B, 2016PhRvL.117f1101S, 2017PDU....15..142C, 2016JCAP...11..036B,2018PDU....22..137C}). 
This is still a controversial question 
(e.g., \cite{2017PhRvD..96l3523A, 2018arXiv180805910B,2016ApJ...824L..31B,2017ApJ...838....8L, 2017PhRvL.119d1102K, 2017PhRvL.118x1101G, 2017PhRvD..95d3534A,2017JCAP...05..017A}),
to which we
pay attention below. PBHs as DM at other mass values ($M\sim 10^{17}$ g, $M\sim 10^{20\div 24}$ g) are still topical (see, e.g., \cite{2017arXiv170402919C} and references given below).
The search for PBHs as the DM is one of the hot-topical issues which gets a new power in the context of PBH cluster models.

There are several models of PBHs formation. PBHs can be formed during the collapses of adiabatic (curvature) density perturbations in relativistic fluid  \cite{p11}. They can be formed also at the early dust-like stages \cite{p25, p40} and rather effectively on stages of a dominance of dissipative superheavy metastable particles owing to a rapid evolution of star-like objects that such particles form \cite{khl1,khl2}. There is also an exciting model of PBHs formation from the baryon charge fluctuations \cite{DolSil93, 2018PhyU...61..115D}.

 Another set of models uses the mechanism of domain walls formation and evolution with the subsequent collapse \cite{p4, p26, dub17}. As we show below, quantum fluctuations of a scalar field near a potential maximum or saddle point during inflation lead to a formation of closed domain walls.  After the inflation is finished, the walls could collapse with black holes in the final state. There is a substantial amount of the inflationary models containing a potential of appropriate shape. The most known examples are the natural inflation \cite{1990PhRvL..65.3233F, 2005JCAP...01..005K} and the hybrid inflation \cite{1994PhRvD..49..748L} (and its supergravity realization \cite{1997PhRvD..56.1841L}). The landscape string theory provides us with a wide class of the potentials with saddle points, see review \cite{2009JHEP...03..149L} and references within.

 In this review, we rely on the models that lead to closed wall formation.
To figure out the idea, let us assume that the potential of a scalar field possesses at least two different vacuum states. In such a situation there are two qualitatively different possibilities to distribute these states in the early Universe. The first possibility implies that space contains approximately equal amounts of both vacuum states, which is typically the case at the usual thermal phase transition. It is supposed that after the phase transition, the two vacuum states are separated by a vacuum wall (kink). This leads to the wall dominated Universe that sure contradicts observational data. The alternative possibility corresponds to the case when the two vacuum states are populated with different probabilities. In this case, islands of a less probable vacuum state appear, surrounded by the sea of another, more probable vacuum state.  The initial distribution of the scalar field which breaks a symmetry between the vacuum states is the essential condition for such an asymmetric final state.

In the framework of our model, the formation of such islands starts at the inflationary stage. Inflationary fluctuations permanently change the value of a scalar field in different space regions leading to the asymmetrical field distribution at the end of inflation. As a result, a distribution of the scalar field is already settled to the end of inflation leading in the future to the islands of one vacuum in the sea of another one. 
The phase transition accompanied by the walls formation takes place only after the end of inflation, deeply in the Friedman–Robertson–Walker (FRW) epoch.
Therefore, the distribution of the scalar field which was formed during the inflation allows the formation of closed walls in the FRW stage. At some instant after crossing the horizon, such walls become causally connected as a whole and begin to contract because of the surface tension. As a result, the energy of a closed wall may be focused within a small volume inside the gravitational radius which is the necessary condition for a black hole creation.

The mechanism of the wall formation is quite general and can be applied to those inflationary models containing extrema of their potential.  Some of them where developed in \cite{PBH_2,2002astro.ph..2505K,Mechanism_BH} and in \cite{2017JPhCS.934a2046G, 2018JCAP...04..042G} where the potentials with saddle points were considered.  It has been revealed there that the PBH mass spectrum and PBH ability depend strongly on the potential parameters and an initial field value, see discussion in Section \ref{sec:detach}. The PBH mass can vary in order of magnitudes. Depending on the chosen parameters, this mechanism can be applied to the description of the dark matter in the form of small BHs or the first quasar formation for massive BHs.  Moreover, the cluster distribution of PBH is realized in a natural way.

The idea of PBH clusterization appears to be rather promising \cite{2018MNRAS.tmp.1331C,2018PDU....19..144G,2015PhRvD..92b3524C}. Indeed, such structures could explain the early quasars observations \cite{2005GrCo...11...99D}, contribute to the dark matter, produce gravitational waves bursts, be a reason for reionization and point-like cosmic gamma-ray sources. 

Another model of the PBH cluster structure was also suggested in the framework of mechanism of black hole formation from strong primordial density inhomogeneities, see \cite{2003ApJ...594L..71A,2006PhRvD..73h3504C,2018arXiv180505912A,2018arXiv180610414D,khl3, 2011PhRvD..84l4031C,  1983ApJ...275..405F},
and discussion in \cite{2018arXiv180505912A}.
The idea of the research is based on the assumption that density peaks in the primordial density fluctuations are clustered more intensively than the ordinary density fluctuations \cite{1984ApJ...284L...9K}.
The latter may imply a deviation from the Poisson distribution of the created PBHs in space.
However, the Poisson effect is also supposed to hold for the PBH clusterization. Besides that, a seed effect, when PBH attracts surrounding matter along with other PBHs, is considered to be the possible independent reason for clusterization. 
The last two effects are used in the series of papers mentioned above for the large-scale structure formation --- galaxies and their clusters (see also \cite{2018MNRAS.478.3756C}).

Heating of the surrounding matter is the inherent property of the domain wall mechanism of PBH (cluster) formation. While collapsing, the domain wall partially transfers its kinetic energy to the ambient matter. It could allow discriminating between the models by observations.

The idea of PBH clustering gives a natural explanation of the origin of massive binary BH. Recent indication \cite{nature25029}  of the possible existence of $\sim 10^4$ black holes with a nearly solar mass within one parsec around galactic center could also be interpreted in the given cluster framework.

There are three aims of this review. Firstly we describe the mechanism that leads to the formation of the PBHs clusters. Secondly, we discuss the space and mass evolution of PBHs in such cluster provided that initial PBHs distribution is assigned. The third aim is to discuss observable effects caused by the PBH clusters. The discussion of cluster evolution and observational implications, conducted in sections \ref{sec:PBH_cluster_form}, \ref{sec:Intern_dynam}, \ref{sec:observ_prop}, being rather general, is illustrated by a specific model described in subsection \ref{specific model}.

\input{Formation/Formation}

\section{Internal dynamics of PBHs inside a cluster}
\label{sec:Intern_dynam}

Previous sections were devoted to the description of the mechanism of PBH formation as the result of walls collapse. The discussion was illustrated by specific model \eqref{Mexica} with the parameters leading to small masses of PBH. The latter could be used in the discussion on the dark matter origin. As is shown in this Section, the same mechanism could be applied for the problem of early quasars. To succeed, the PBH should be made more massive.

The clusters with massive PBHs can be obtained in the framework of not only the same mechanism but even with the same specific model \eqref{Mexica}. The only thing that should be altered is the model parameters. We demonstrate this in our first paper \cite{dub17}. In this Section, we alter significantly the parameter $\Lambda$ only~--- from $\Lambda=0.05$ in previous Sections to $\Lambda=5$ in this Section.

It will be shown that the virialized cluster contains a central supermassive black hole that can serve as the seed for an early quasar. The origin of the dark matter is not discussed in this Section. The model for explaining both the dark matter and the early quasars on the basis of the same approach is elaborated.

\subsection{Cluster detachment from the Hubble flow\label{sec:detach}}

\input{Detachment/Detachment}

\subsection{Internal dynamics after detachment \label{InternalDynam}}

The study of the gravitational dynamics of PBHs clusters after detachment reduces to the problem of $N$-body simulation that can be solved only numerically. For these purposes we have used the \texttt{NBODY6++} code \cite{2015MNRAS.450.4070W}, aimed at the simulation of globular stars clusters. We have made some changes  to the program: all bodies have been deprived of their stellar characteristics (luminosity, types of star and their evolution, metallicity, etc.), objects' <<sizes>> have been redefined according to their gravitational radius ($r_i={r_g}_i=2GM_i/c^2$, where $r_g$~ is the gravitational radius),  merging mechanism has been added.

\begin{figure}[t]
 \centering
 \includegraphics[width=0.6\linewidth]{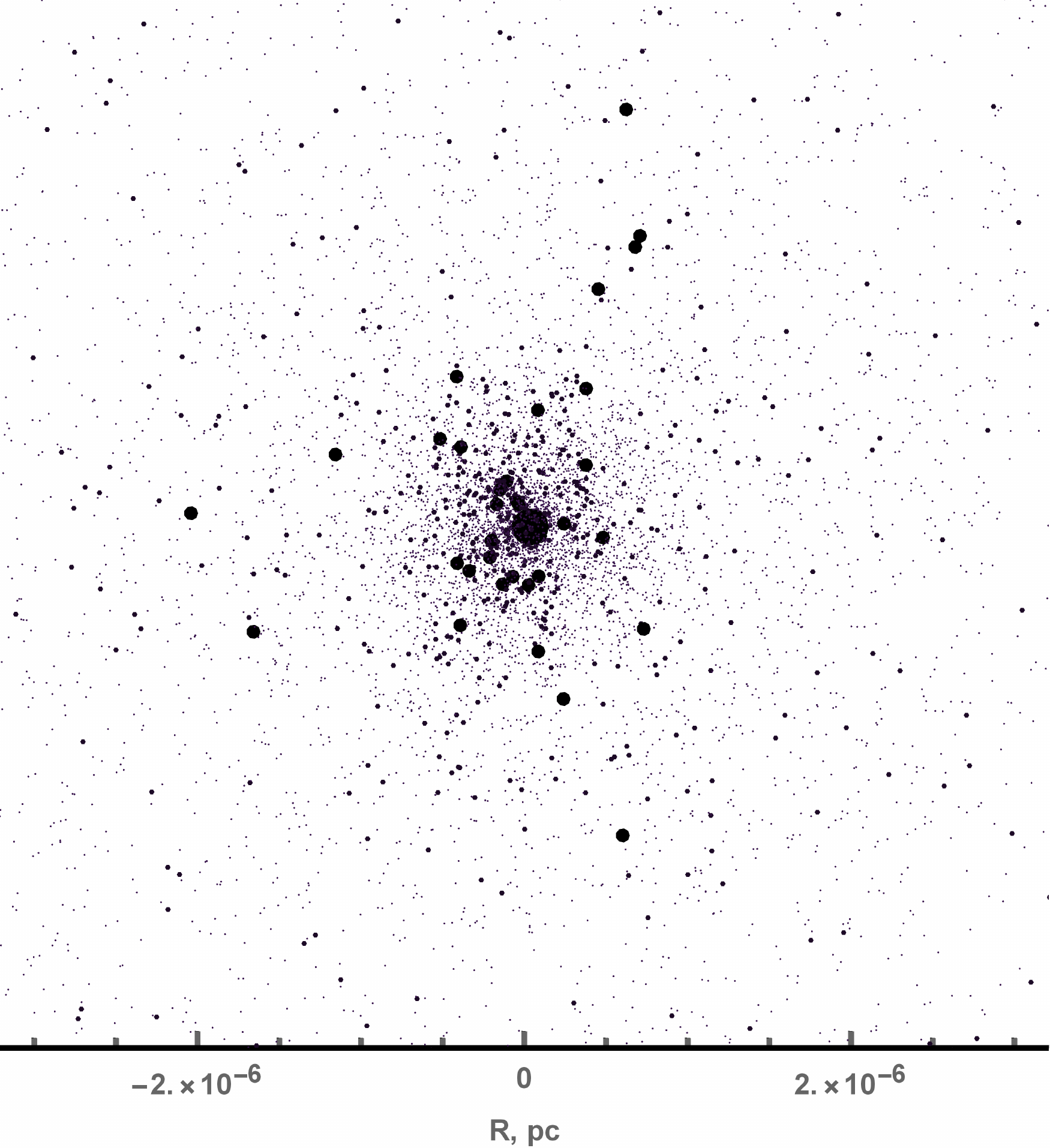}
 \caption{The typical spatial black holes distribution within cluster, the result of numerical simulations. Black holes <<sizes>> show its mass distribution but do not correspond to real ones. 
 }
 \label{fig:Space_distrib_detach}
\end{figure}

Here we present the first results of the $N$-body simulation of the PBHs cluster evolution. In the numerical simulation, we included only BHs with $M>0.1 M_\odot$ due to limited computer resources. Such BHs define the main inhomogeneities in the density structure (see Fig.~\ref{fig:N_r_detach} and \ref{fig:Space_distrib_detach}). 
Further, PBHs with other masses will be taken into account as far as possible since they can give noticeable dynamical friction effects and modify mass distribution due to changing merger process probability.

Let us consider a typical example of simulation with $N=10^4$ initial black holes in the cluster and the mass distribution
\begin{equation}
    \frac{dN}{dM} \propto \frac{1}{M_\odot} \left(\frac{M}{M_\odot}\right)^\alpha.
    \label{dNdM1}
\end{equation}
The spatial distribution is chosen in the form
\begin{equation}
    \rho(r)\propto
    \begin{cases}
        r^{-\beta}, & r\leq R,\\
        0, & r> R.
    \end{cases}
    \label{rho_r}
\end{equation}
Here $R=1$~pc is the cluster size. The initial velocities were chosen according to the Maxwell distribution with the virial velocity $v_0 \sim 1$~km/s. We assume the most massive black hole to be in the center of the cluster with zero initial velocity.

The PBH distribution in Fig.~\ref{fig:N_r_detach} is approximated by expressions \eqref{dNdM1}--\eqref{rho_r} with $\alpha=-2.4$ and $\beta=2$ for the central region of cluster in the mass range $10^{-1}\div10^{2} M_\odot$ \cite{2017JPhCS.934a2040N}.
The power $\alpha=-(2\div 3)$ in \eqref{dNdM1} is typical of the considered model with simple potential \eqref{Mexica}. 
The mass distribution within the cluster slightly changes due to mergers of PBHs and their escape from the cluster.

\begin{figure}[t]
 \begin{center}
  \subfigure[]{
    \includegraphics[width=0.47\textwidth]{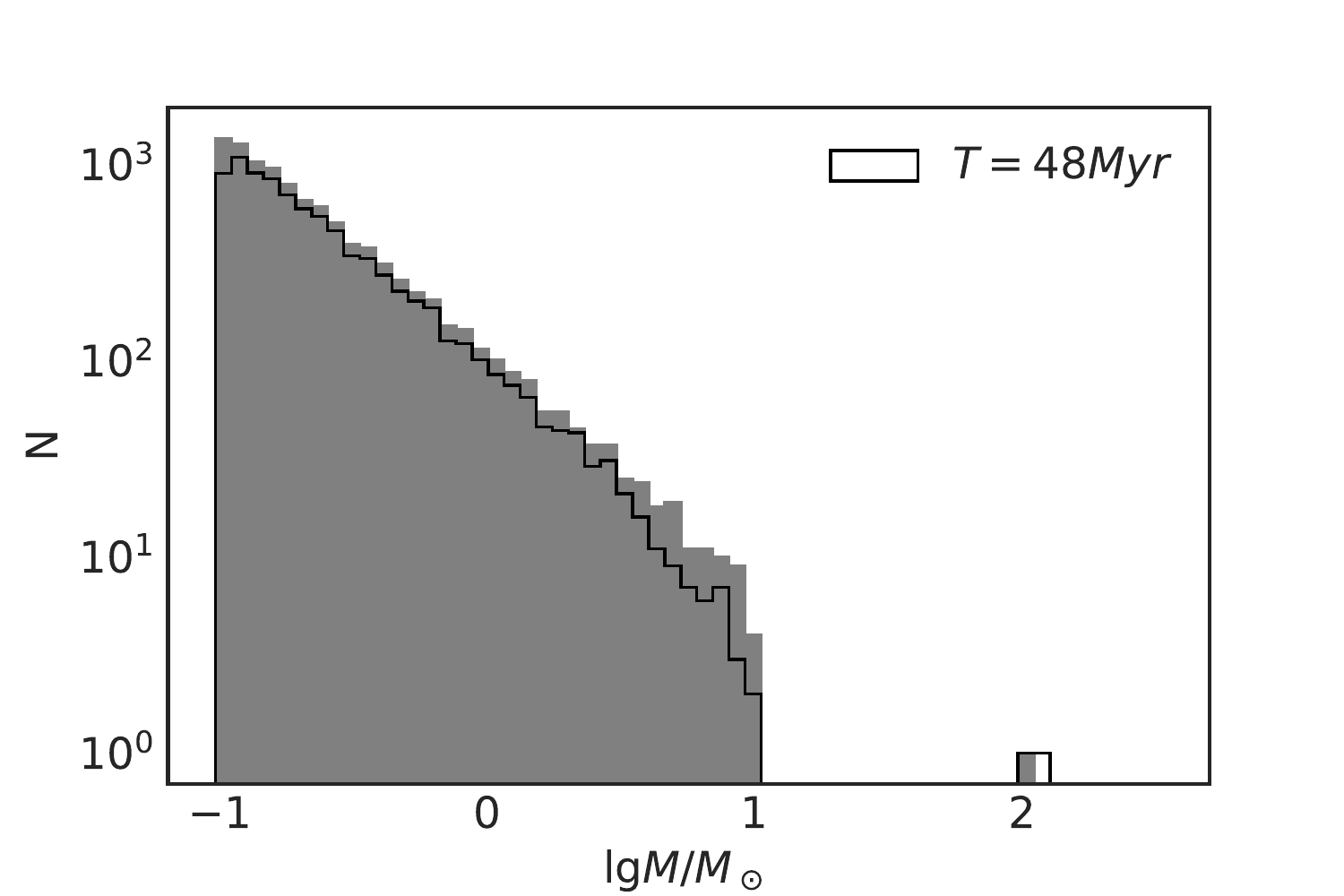}
   \label{fig:mass}
  }
  \subfigure[]{
   \includegraphics[width=0.47\textwidth]{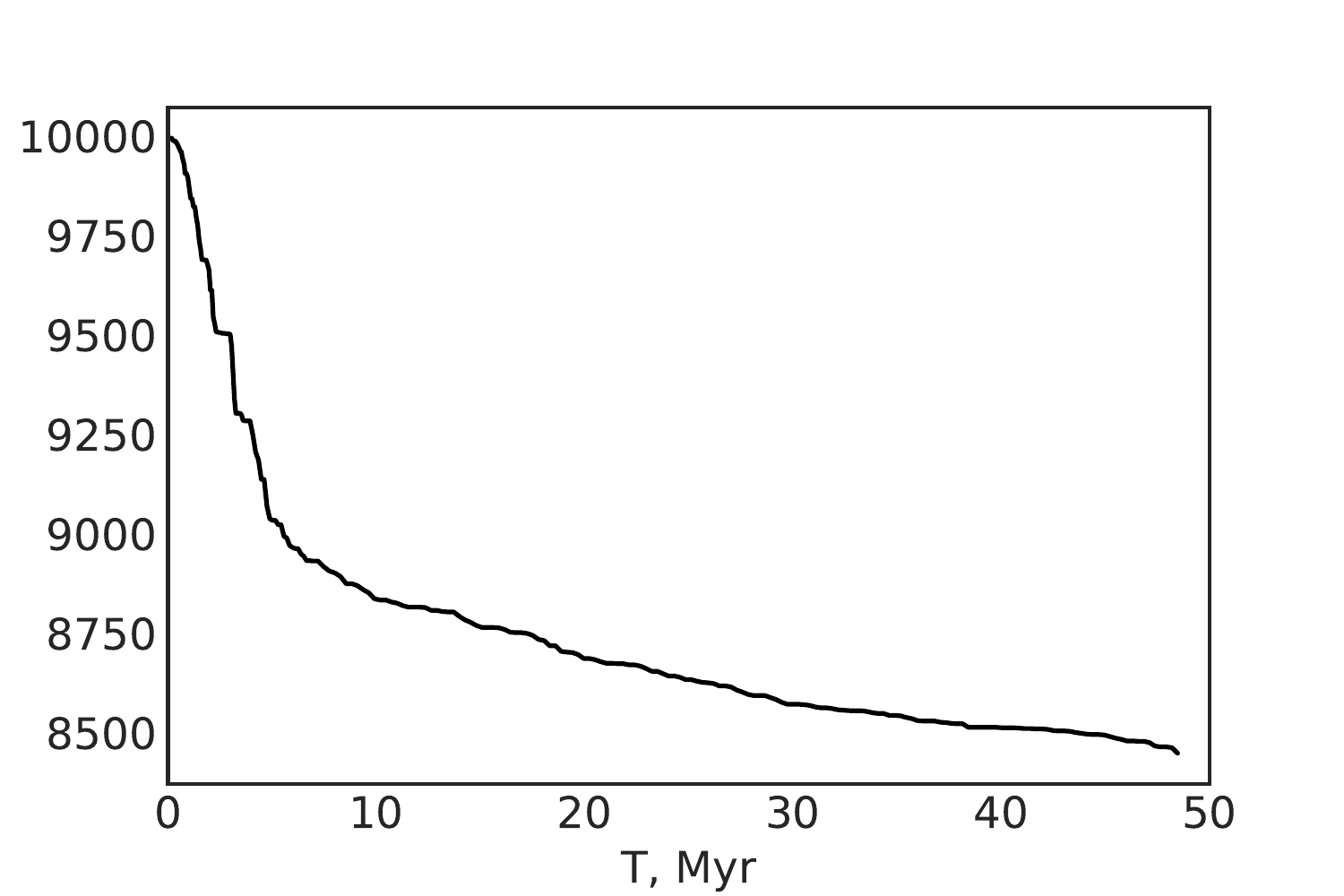}
   \label{fig:num}
  }
 \end{center}
 \caption{Dynamics of PBHs cluster with the total mass 3800~$M_\odot$ and mass distribution \eqref{dNdM1} consisting of the $N=10^4$ bodies starting from $z\sim10^4$ until the moment $z\sim40$.
 (a): Сhange of the mass spectrum from the initial one (filled gray area) to the final (solid black line). (b): Сhange of the total number of PBHs in the cluster due to merging and escape during the same period}.
 \label{fig:Mass_Num_Evol}
\end{figure}

The results are presented in Fig.~\ref{fig:Mass_Num_Evol}. One can see that cluster as a whole does not change its structure significantly: the mass spectral index is not changed essentially ($\alpha_\text{f}\approx -2.4$), the number of particles and the total cluster mass is reduced by $\sim15\%$ . The cluster is not destroyed and remainsa gravitationally bound system by the moment $z\sim20\div40$, when, according to estimates, the accretion starts influencing considerably the growth of the most massive black holes (see \cite{2018Natur.553..473B}). The numerical calculations are to be further developed in this field. Existing $N$-body simulation codes like \texttt{NBODY6++} should be adapted for these purposes. In particular, modeling the dynamics of massive black holes is still quite problematic.

Notice that the total mass of PBH cluster is enough to form the early quasar (at $z>6$). The problem of the early quasar origin is widely discussed. The simulations show that early quasars formation is possible in the case if by the moment $z\sim 20\div40$ there had already existed their <<seeds>>~--- either <<abnormal>> baryonic objects that further collapse into black holes or black holes themselves with masses $10^{3\div 5}M_\odot$ (such masses are required to get quasars with mass $M\sim10^{9}M_\odot$, which are mostly observed at $z\sim6$ \cite{2018Natur.553..473B}). However, the formation of massive <<seeds>> at such high redshifts requires additional explanation. Our analysis shows that we can offer one more scenario for the natural formation of these <<seeds>>.

\section{Observable properties}
\label{sec:observ_prop}

As already mentioned, the PBH clusters could be a clue to the solution of different problems in cosmology and astrophysics, identified from different observations. These are dark matter (DM) problem, supermassive black holes (SMBH) origin, black hole mergers registered by LIGO/Virgo, also the origin of unidentified cosmic gamma-ray sources, reionization of the Universe and maybe others. Here we list them as possibly applied to the PBH cluster model.

\subsection{Supermassive black holes}

The fact that each galaxy contains a supermassive black hole is almost no longer in doubt (see, e.g., the latest results of Chandra experiment \cite{2017ApJS..228....2L}). Standard mechanism of SMBH formation requires a quite long time to reach so large black hole mass and is hardly consistent with SMBH existence at the redshift $z\gtrsim 4\div 5$ \cite{2015MNRAS.451.1892A,2018Natur.553..473B}. However now a few dozens of quasars are observed at higher redshift ($z\sim 7$), which are considered to radiate due to intense accretion onto supermassive (as a rule $\sim 10^9 \,M_{\odot}$) black hole. Especially one can note an observation of quasar at $z\approx 6.3$ with the mass $12\times 10^{9} M_{\odot}$ \cite{2015Natur.518..512W}.

In PBH cluster mechanism, such quasars can be formed as a result of accretion of seed PBH \cite{2005GrCo...11...99D,2007arXiv0709.0070D,2010RAA....10..495K,2012Sci...337..544V,2018PhyU...61..115D,2018arXiv180101095K}. The mass of the seed PBH is defined by cluster's formation considered above and its subsequent evolution.

As was said just above, getting PBH of the mass $M\sim 10^3M_{\odot}$ at the redshift $z\sim 20-40$ could give rise to SMBH of $M\sim 10^9M_{\odot}$ at $z\sim 7$ due to accretion in critical regime (see, e.g., \cite{2018Natur.553..473B,2018arXiv180101095K}).

A hypothesis on the primordial origin of SMBH is suggested to be probed in the measurement of 21 cm hydrogen line \cite{2018JCAP...05..017B}. The observational effect should be induced by the accretion of surrounding matter onto PBH during the Dark Age epoch.

Existent correlation between X-ray and infrared backgrounds can be treated in favour of massive PBH as a seed of star formation at high redshift ($z\sim 20$) \cite{2018PDU....22..137C}.

If a cluster contains many PBHs of intermediate mass, it can give the opportunity to check or limit the model with the help of LIGO and LISA \cite{1980A&A....89....6C, 2017arXiv170510361K}. In the recent work \cite{2018PhRvD..98b3543B} such attempt is undertaken concerning the clusters, which is commented below.

The intriguing possibility which would be worth considering in the framework of the cluster model \cite{quasars_Laletin} is the effect of a spin alignment of the large groups of quasars which was recently observed for two quasar groups (at $z\approx 1.3$) \cite{2016A&A...590A..53P}.

\subsection{PBH clusters and observational limits on PBH dark matter}

If PBHs are formed in an appropriate amount, they can play the role of the dark matter (DM) in the Universe \cite{p23}. However, the <<fine tuning>> is required to avoid the $\Omega_{\rm PBH}\ll1$ or $\Omega_{\rm PBH}\gg1$ situations.

\begin{figure}[t]
    \centering
\includegraphics[width=0.7\textwidth]{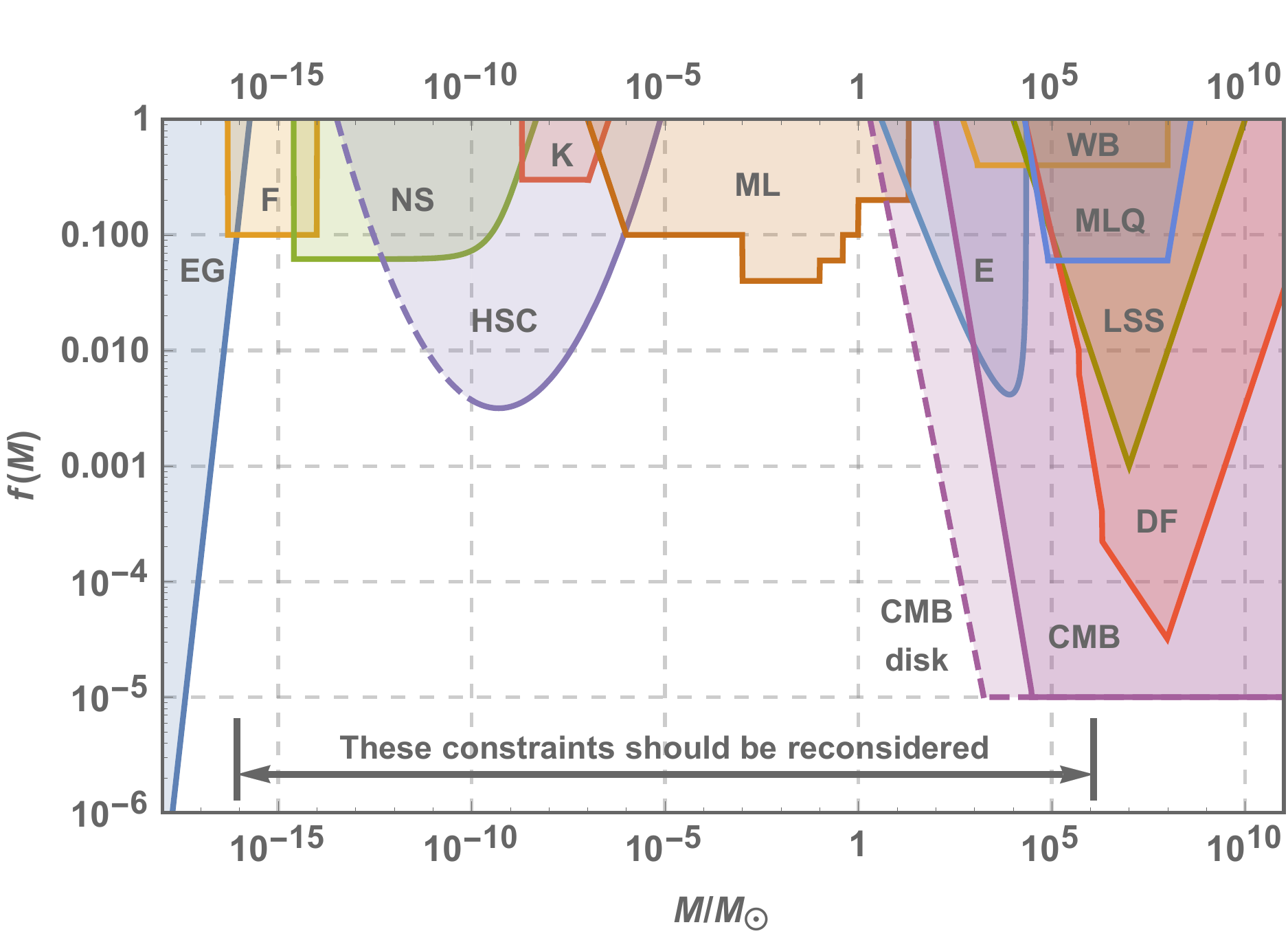}
\caption{Existing constraints on relative contribution of the uniform PBHs distribution into dark matter density with monochromatic mass distribution.
The most robust constraints in their regions are basically depicted following the notation of \cite{2016PhRvD..94h3504C}: from extragalactic gamma-ray background (EG) observation, femtolensing (F), neutron star destruction (NS), searches for gravitational microlensing events by Subaru Hyper Suprime-Cam (HSC), Kepler satellite (K), MACHO (ML), quasars millilensing (MLQ), wide binary (WB) and star cluster in Eridanus dwarf galaxy (E) destruction, CMB distortion due to accretion effects (WMAP), effects of dynamical friction in our Galaxy (DF) and large-scale structure (LSS). 
CMB constraint was taken for (more minimal) case of spherically symmetric accretion \cite{2017PhRvD..95d3534A} in comparison with the less minimal (``CMB disk'') one \cite{2016PhRvD..94h3504C} (shown by dashed line). Double-headed arrow shows region where the constraints are to be re-considered for PBH clusters.}
\label{Omega}
\end{figure}

As a dominant form of DM, PBHs have been proposed a long time ago; special activity started from the first results on the MACHO search \cite{1997ApJ...486..697A}. By now there have appeared many constraints on such candidate in the very different mass ranges (see Fig.\ref{Omega}). 

A comprehensive review on that can be found in \cite{2016PhRvD..94h3504C}; there is one more appeared recently \cite{2018CQGra..35f3001S}. Since the first review (mid-2016), several new constraints had time to appear. The search for microlensing events of stars in M31 with HSC telescope \cite{2017arXiv170102151N} constrains PBHs with mass $10^{20}\div 10^{27}$~g (interval should be cut from below with $10^{23}$~g because of violation of geometric optics approximation \cite{2018arXiv180206785K}). Also, additional analysis of ionization and thermal history of the Universe through the data on cosmic microwave background (CMB) anisotropy allows updating constraint on PBH abundance in different mass ranges. In the light mass range, the analysis provides constraints in a narrow range around $M\sim 10^{15}$~g a little bit stronger than existed
\cite{2017JCAP...03..043P,2017PhRvD..95h3006C}, which originates from CMB distortion by Hawking evaporation. Re-consideration of accretion effects taking into account that they may proceed in a disk regime (not spherically symmetric one) may lead to ruling out PBHs as DM for $M\gtrsim M_{\odot}$ (plus-minus an order of magnitude depending on the velocity parameters) \cite{2017PhRvD..96h3524P} from CMB data. Also higher mass interval of PBH density is claimed to be constrained by the data on Super Novae (SN) gravitational lensing (for $M\gtrsim 10^{-2}\, M_{\odot}$) \cite{2017arXiv171202240Z}, and by possible heat deposition in baryons limited with 21-cm absorption line observation by EDGES experiment at $z\approx 17$ (for $M\sim 10\, M_{\odot}$) \cite{2018arXiv180309697H}.
However, SN constraint is criticized \cite{2017arXiv171206574G}, and EDGES data are to be confirmed. There is also some dispute about microlensing constraints for $M\sim 10 M_{\odot}$ \cite{PhysRevD.96.043020, 2018MNRAS.tmp.1331C}. However, they may be not so crucial in this mass range. Though some new constraint at $M\sim 10^{-6}\div 10^{-3}M_{\odot}$ appears from OGLE data on microlensing events \cite{2019arXiv190107120N}, it allows interpretation as the indication of the Earth mass PBH with percent fraction in DM. There appear new constraints from BH merger events registered by LIGO. Work \cite{2018JCAP...10..043B} puts some constraint on PBH amount at $M\sim 10 M_{\odot}$ considering the contribution of gravitational waves from BH mergers to relativistic matter component 
in the Universe; we comment it together with the constraint from \cite{2018arXiv180805910B} for the cluster below.

Imposition of all the restrictions does not seem to leave a chance for PBH to be a dominant DM candidate. However, the current situation is more complicated. Many works were arguing that constraints coming from CMB, star cluster disruption in Eridanus dwarf galaxy and microlensing effects in the mass range $\sim 1\div 100 M_{\odot}$ for uniformly distributed PBHs can be weaker. Therefore, both dark matter and gravitational waves from BH merger could be explained by PBHs (see, e.g., \cite{2018MNRAS.478.3756C, 2018PDU....22..137C, 2019arXiv190107803C}).  Also, PBHs of the mass within two gaps around $10^{17}$ g  and $10^{20\div 24}$ g are so far being considered as DM candidates.

Also, there are suggested possible constraints to be obtained by LISA in future coming from observation of gravitational waves signal from PBHs sinking to SMBH in our Galaxy \cite{2017arXiv170510361K, 2017arXiv170903500G} (in the ranges $10^{20}\div 10^{24}$ g and $10^{-2}\div 10^{-1} M_{\odot}$), and pulsar timing effects (delay of signal from pulsar) induced by PBH moving along the line of sight \cite{2017PhRvD..95b3002S} (in the PBH mass interval $1\div 10^{3} M_{\odot}$). EDGE experiment is suggested to put a constraint on PBH of masses $10^{-5}\div 10 M_{\odot}$ \cite{2018JCAP...11..041G} by investigating density inhomogeneity through observation of 21 cm hydrogen line as it was done with Ly-$\alpha$ clouds \cite{2003ApJ...594L..71A} . These new possible constraints may overlap some still not yet (fully) forbidden mass ranges. 

However, one should note that all the constraints mentioned above have been obtained basically for monochromatic PBH mass distribution. As it was shown above it is not the case of PBH cluster as well as of many other PBH models \cite{2016PhRvD..94h3504C}. In the papers \cite{2017PhRvD..95h3508K,2016PhRvD..94f3530G,2018CQGra..35f3001S,2017PhRvD..96b3514C} the constraints are tried to be generalized for extended mass spectrum (everywhere a lognormal distribution is considered and also power-law and exponential ones in the last reference). No weakening restriction was virtually obtained. 

But not every type of constraint can be generalized <<universally>>: i.e., as if PBHs with different mass values contribute to the same observably constrained value (say, number of lensing events, or upper limit on the number of destroyed white dwarfs or neutron stars, etc.). For instance, constraint coming from gamma-ray background relevant for PBH with the mass $M\sim 10^{15}\div 10^{17}$~g cannot be reduced to the <<universal>> generalization, since PBHs with different mass contribute to different parts of energy $\gamma$-spectrum. As was shown in \cite{2017IJMPD..2650102B}, with power-law mass distribution 
\begin{equation}
\frac{dN}{dM}\propto M^{\alpha},\;\;M_{\min}<M<10^{17}\text{ g}
\label{power}
\end{equation}
one can gather all DM in the form of PBHs. The upper limit of $M$ in \eqref{power} is defined by a mass value starting from which constraint from a femtolensing comes into force.

\begin{figure}[t]
    \centering
    \includegraphics[width=0.6\textwidth]{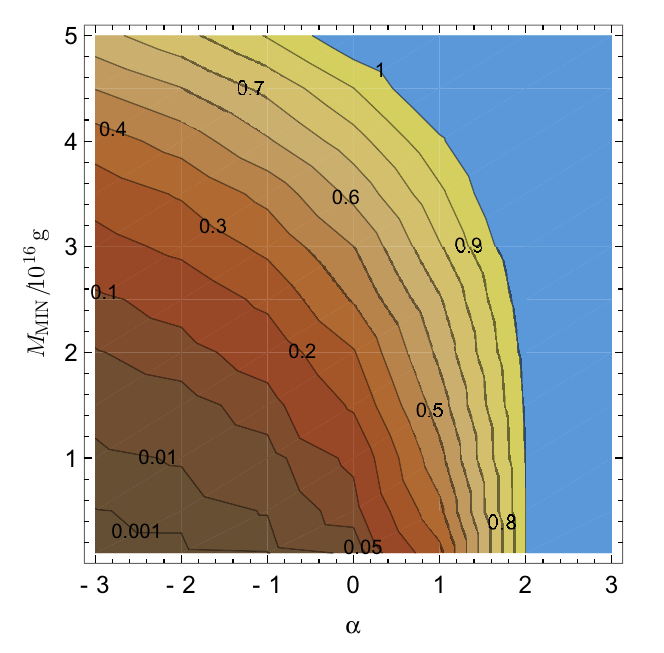}
\caption{Relative contribution of PBH with $M\sim 10^{16}$ g and power-law mass distribution into DM in dependence on degree of the power-law $\alpha$ and lower cut in mass $M_{\rm min}$ (see \eqref{power}).}
\label{Omega_15}
\end{figure}

We quote figure here \ref{Omega_15}, which shows a quite big parameter range (blue region in the coloured figure) where PBHs can substantially contribute to DM and simultaneously to reionization of the Universe (see Fig.~3 left of \cite{2017IJMPD..2650102B} and discussion below in subsection \ref{reion}). The PBH as DM with the mass around the same value is also considered in other works, e.g., \cite{2018arXiv180206785K,2018PhRvL.120l1301E}. Contribution to DM can also be maximized with several delta-functions (peaks) at different masses of PBH mass spectrum \cite{2017IJMPD..2650102B,2018JCAP...04..007L}.

In the cluster model, a mass distribution can be very different depending on the model parameters (mostly of a scalar field potential). The case considered above gives nearly decreasing power-law from PBH mass ($\alpha\approx -(2\div 3)$). However, one can get growing power law as well as distribution with several peaks at different mass values. However, it is not the only opportunity to change or even avoid PBH abundance limitations in case of a cluster.

Almost all the existing constraints on PBH abundance with monochromatic mass distribution have to be reconsidered in the case of space and mass distributed PBH clusters. For instance, all constraints from femto-/micro-/milli- lensing can be irrelevant since it is a cluster as a whole that should give a lensing effect, while its size may exceed the Einstein radius. Besides, the probability of lensing for smaller PBHs can be suppressed because of their inhomogeneous (clustered) distribution in space with effectively diminished number density to the cluster's number density. However one should take into account that there can be an unclustered component. The latter could consist of those PBH that escaped by clusters, and the clusters themselves could be disrupted in Galaxy as well.

The investigation of the lensing effect for the cluster which size exceeds the Einstein radius is underway now by our group.
The opposite case is considered in \cite{2018MNRAS.tmp.1331C,2018PDU....19..144G}. They take the cluster model \cite{2003ApJ...594L..71A,2006PhRvD..73h3504C} with the lognormal mass distribution of PBHs inside the cluster. Some increase of total PBH abundance, in this case, can be reached due to an increase of PBH amount inside the cluster which shifts and broadens the mass distribution of the clusters what may make limitations a little more flexible.

As to dynamical constraints, they should also be reconsidered. These are disruption of white dwarfs, neutron stars, wide binary systems, star clusters, etc. Constraint coming from star cluster disruption in dwarf galaxy looks like one of the strongest and reliable for the PBH clusters. The constraint could relate to the PBHs of masses around $\sim 10 M_{\odot}$ inside the cluster. 

Also in similar mass range, existing data on gravitational waves can constrain the cluster abundance, as claimed in the recent work \cite{2018PhRvD..98b3543B}. However, the process of PBHs pairs formation considered there seems to be more relevant at early stages (before virialization and decoupling from Hubble flow) of cluster evolution. Also, one should note that these constraints, if they hold, can be ineffective for the PBH cluster model considered here. PBHs of the mass $\sim 10 M_{\odot}$ are not expected to give a dominant contribution to DM density provided by falling power-law mass spectrum \eqref{dNdM},\eqref{power}. If we take $\alpha = -2.4$, as obtained above, and the mass range
\begin{equation}
0.5\times10^{15}\text{ g}<M<100 M_{\odot},
\label{mrange}
\end{equation}
the relative contribution of PBHs with $M>M_{\odot}$ to their total density is less than $10^{-7}$. Note that the range \eqref{mrange} is different from the one above. Nonetheless, it could be got so by customizing initial parameters. However, it would strongly complicate the simulation of the cluster.

As to the constraint from CMB, based on consideration of accretion by PBH with mass $\gtrsim M_{\odot}$ (most robust constraint for most massive PBHs), it can be weakened for clusters due to a motion of PBH inside them, changing the regime of accretion \cite{2018PDU....19..144G}. At the same time, if one takes the very massive PBHs, their amount is not supposed to be large to explain seeding SMBH in galaxy centers.

Finally, the majority of constraints considered and obtained in \cite{2016PhRvD..94h3504C} for the monochromatic PBH mass distribution is worth being re-considered for PBH clusters with extended mass distribution both for PBHs inside the cluster and for clusters themselves. In Fig. \ref{Omega} we point out the PBH mass interval of constraints obtained for monochromatic mass distribution which is to be re-considered for clusters. We suppose that this interval can be significant. 

There is another advantage of considering the (clustered) PBHs as the dark matter. As was noticed in  \cite{2018PDU....22..137C}, their gravitational interaction allows circumventing the problems of standard non-interacting CDM scenario like cusp crisis, dwarf galaxies overproduction, etc.

\subsection{Gravitational waves from PBH coalescence}

By now ten BH merger events have been detected by LIGO/Virgo \cite{2018arXiv181112907T}; the first five announced \cite{2016PhRvL.116f1102A,2016PhRvL.116x1103A,2017PhRvL.118v1101A,2017PhRvL.119n1101A,2017ApJ...851L..35A} are given in the Table ~\ref{table}. It seems from viewpoint of standard astrophysical origin to be rather unnatural, that they are pairs of heavy black holes ($\gtrsim 10 M_{\odot}$) with low initial (effective inspiral) spins. 
More arguments \textit{contra} BH of star origin and \textit{pro} primordial one for given events are collected in \cite{2018PDU....22..137C}.

\begin{table}[t]
\caption{\label{tab:canonsummary}The first five announced BH merger events detected by LIGO/Virgo.}
\renewcommand{\arraystretch}{1.8}
\renewcommand{\tabcolsep}{0.4cm}
\begin{center}
\begin{tabular}{ |c|c|c|c| }
\hline
Event & BH$_{1}$ mass, $M_{\odot}$ & BH$_{2}$ mass, $M_{\odot}$& Effective inspiral spin $\chi_{eff}$  \\
\hline
GW170608 \cite{2017ApJ...851L..35A} & $12^{+7}_{-2}$ & $7^{+2}_{-2}$ & $0.07^{+0.23}_{-0.09}$ \\
\hline
GW170104 \cite{2017PhRvL.118v1101A} & $31.2^{+8.4}_{-6.0}$ & $19.4^{+5.3}_{-5.9}$ & $-0.12^{+0.21}_{-0.30}$ \\
\hline
GW170814 \cite{2017PhRvL.119n1101A} & $30.5^{+5.7}_{-3.0}$ & $25.3^{+2.8}_{-4.2}$ & $0.06^{+0.12}_{-0.12}$ \\
\hline
GW151226 \cite{2016PhRvL.116x1103A} & $14.2^{+8.3}_{-3.7}$ & $7.5^{+2.3}_{-2.3}$ & $0.21^{+0.2}_{-0.1}$ \\
\hline
GW150914 \cite{2016PhRvL.116f1102A} & $36^{+5}_{-4}$ & $29^{+4}_{-4}$ & $-0.06^{+0.14}_{-0.14}$ \\
\hline
\end{tabular}
\end{center}
\label{table}
\end{table}

So, there are already serious attempts (starting from \cite{1997ApJ...487L.139N}) to explain such binary BHs existence in the framework of PBH models (e.g., \cite{2018arXiv180509034K,2018arXiv180507757A,2018arXiv180505912A,2018arXiv180206785K,2018arXiv180110327C,2018CQGra..35f3001S,2018arXiv180107360W,2017arXiv171109702G,2017JCAP...09..037R,2016JCAP...11..036B,2018PhyU...61..115D,p26,dub17,2018PDU....22..137C}). Those from them of cluster \cite{2003ApJ...594L..71A,2006PhRvD..73h3504C,2018arXiv180505912A,p26,dub17,2018PDU....22..137C} inevitably predict such binaries, so having good prospects to be probed with (to be) appeared gravitational tools LIGO/Virgo, LISA (see references above). Though some (aforementioned) attempts to put constraints from LIGO/Virgo data on the abundance of clustered or binary PBHs already starting to appear \cite{2017PhRvD..96l3523A, 2018arXiv180805910B, 2018JCAP...10..043B}.

Although it is possible that binary BHs with masses of $\sim 30 M_{\odot}$ can be formed after deaths of Population III stars with low metallicity. But there still exist large uncertainties in theoretical predictions of the merger rate by a factor of O(100) due to unknown astrophysical parameters \cite{2016MNRAS.456.1093K,2016Natur.534..512B}. Such possible origin could be confirmed by observation of an electromagnetic signal, but no electromagnetic counterparts were observed for BH merger events \cite{Zakharov.icppa-2018}. It is worth mentioning once again counter-arguments for star origin given in \cite{2018PDU....22..137C}.

The rate of PBHs collisions in clusters was calculated in \cite{DokEroRub09} for the planned LISA (Laser Interferometer Space Antenna) instrument sensitivity curve in the frequencies range $10^{-4}-1$~Hz. Although the LIGO/Virgo interferometers fulfilled the first detection of gravitational waves, the primary approach of \cite{DokEroRub09} is applicable to LIGO/Virgo too, but the preferred masses for detection move to lower values due to the smaller working frequencies of the ground-based detectors. We plan to perform the similar calculations for LIGO/Virgo in future works, but here we outline the general ideas and present the numerical results, obtained in \cite{DokEroRub09} for LISA. These predictions can be used if the LISA project will finally be realized. 

The initial mass spectrum of PBHs in the cluster is supposed to be the same as in the paper \cite{lisa11}.  According to this work, the distribution just after the cluster virialization can be fitted as 
\begin{equation}
\frac{dn}{dM}= 1.6\times10^{3}
\left(\frac{r}{1\mbox{~pc}}\right)^{-3}\left(\frac{M}{M_{\odot}}
\right)^{-2}\mbox{$M_{\odot}^{-1}$pc$^{-3}$}. \label{bhfit1}
\end{equation}
The total mass of the PBHs prevails that of the dark matter at the distances $r<1.6$~pc from the center, and we restrict our analysis to this region. 
It should be taken into account that two-body relaxation in the cluster leads to the contraction of the inner region and the merging of the inner shells with a central black hole. For each moment, we find the radius of the full relaxation and exclude its interior from the calculations.
Then we calculate the rate and the distribution of the PBHs collisions. The cross-section for two PBHs capture and subsequent merging is  \cite{lisa12}
\begin{equation}
\sigma_{\rm mer}=2\pi\left(\frac{85\pi}{6\sqrt{2}}\right)^{2/7}
\frac{G^2(M_0+M)^{10/7}M_0^{2/7}M^{2/7}} {c^{10/7}v_{\rm
rel}^{18/7}}, \label{m1m2}
\end{equation}
where $M_0$, $M$ are the masses of the PBHs, and $v_{\rm rel}$ is their relative velocity which is assumed to be of the order of virial velocities in the cluster center. With this cross-section, the rate of the PBHs collisions in the cluster can be calculated as
\begin{equation}
\label{ddN} d\dot N=\sigma_{\rm mer}v_{\rm rel}\,dn,
\end{equation}
where $dn$ is the PBHs number density with masses in the interval $M\div M+dM$.  It was found in \cite{DokEroRub09} that the main signal (for LISA) comes from the collisions of PBHs in the cluster with the central most massive black hole. Note that in the case of LIGO/Virgo the mutual collisions between smaller black holes in the cluster would be more important. The signal-to-noise ratio can be calculated as \cite{lisa15} 
\begin{equation}
 \rho_{\rm SN}(z,M)^2=
 4\int\frac{|\tilde h(f)|^2\,df}{S_h(f)},
 \label{sn}
\end{equation}
where $S_h(f)$ is the noise spectrum of the detector. The function $\tilde h(f)$ represents the spectrum of the signal. It was calculated  in \cite{lisa15} for the local population of sources ($z\ll1$) and was obviously generalized by \cite{DokEroRub09} for arbitrary red-shifts $z$. For each PBH's mass $M$ there is a distance threshold -- maximum redshift $z$ from which it can be detected. We divide the full mass interval into several smaller mass ranges and calculate the gravitational bursts rate for each of the intervals. We also suppose that the merging of the intermediate-mass black holes in the cluster finally produces the supermassive black holes in the typical galaxies. This assumption allows one to fix the number density of the clusters in the universe. Abolishing this assumption one obtain different normalization of the events rate. 

\begin{figure}[t]
\begin{center}
\includegraphics[width=0.7\textwidth,trim={3cm 0cm 2cm 20cm},clip]{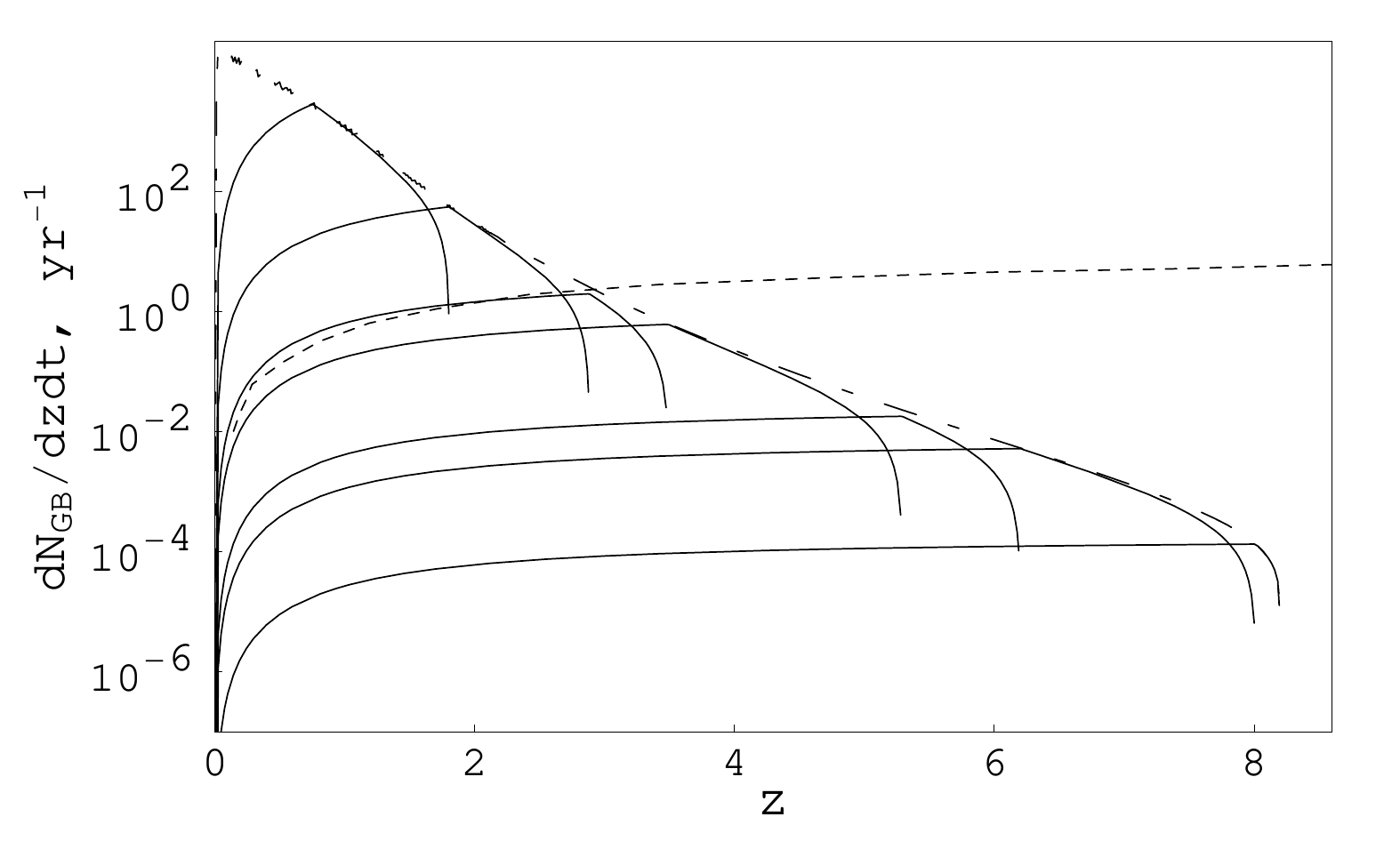}
\end{center}
\caption{Gravitational bursts distribution by sources (PBHs clusters) red-shifts $z$.
The solid curves correspond to the total merger rates for the PBHs in
the mass intervals $M=10^2-5\times10^2M_{\odot}$,
$5\times10^2-10^3M_{\odot}$, $10^3-5\times10^3M_{\odot}$,
$5\times10^3-10^4M_{\odot}$, $10^4-5\times10^4M_{\odot}$, and
$5\times10^4-10^5M_{\odot}$ (from up to down). The breaks on the curves correspond to the maximum red-shifts of LISA detection in these intervals.  Envelope bar-dotted curve shows the total rate. For comparison
the dashed line shows the result of \cite{lisa17}.}
\label{figlisa1}
\end{figure}

The results of the calculations are presented in Fig.~\ref{figlisa1}. One can see that the detection of the gravitational bursts by LISA would be possible during the reasonable time of observations. The similar study for LIGO/Virgo is desirable and is planned soon. The dependence of the rate on the red-shift $z$ in our model differs sufficiently from the one obtained in the model \cite{lisa17}, where astrophysical black holes, formed from the collapses of the gas clouds, were considered. This specific redshift dependence of the merger rate is determined by the early beginning of the two-body relaxation in the cluster centers. It led to the central density increase and the merger rate amplification. The difference in the redshift dependence could help to discriminate between models of black holes formation.

Notice that there is a contribution to the gravitational waves from the very light PBHs which can be experimentally accessible in future \cite{2018arXiv180805910B,2018arXiv181012218B}.

\subsection{PBH clusters as point-like gamma-ray sources}

Due to weak accretion of surrounding matter, observation of BHs with masses less than $10^4\, M_{\odot}$ can be quite difficult.
An alternative method for the BHs detection through the Hawking’s radiation \cite{Haw75} is useful only for BHs with low masses that are situated near the Earth.  From \cite{BHClust2011,2011APh....35...28B} one can conclude that situation can be different for PBH clusters with a large number of small BHs (with masses $\lesssim 10^{15}$ g).  The integral Hawking’s radiation of such a cluster could then be sufficient for detection by modern gamma-ray telescopes on the Earth (in the near-earth space). Now there is a big database on unidentified point-like cosmic gamma-ray sources (PGRS) obtained by Fermi-LAT \cite{2010ApJS..188..405A} which has more than 500 such sources.

The most of $\gamma$-ray sources (more than 60\% of their total amount) are associated with active galactic nuclei, $\gamma$-ray pulsars and some other Galactic sources. However, there is an essential part of PGRS without counterparts found in other wavelengths and thus non-identified. They could have a conventional origin and can not be identified because of insufficient angular resolution  
but also could be of unknown physical origin, which attracts so far special interest (e.g., \cite{2014arXiv1412.1930D, 2016arXiv160300231S,2018arXiv180203764T}).

We made estimations for specific case of cluster model \cite{BHClust2011,2011APh....35...28B} following \cite{lisa11}. Fractal-like PBH cluster structure is supposed to be formed around the most massive PBH being the progenitor of SMBH. It, in turn, is assumed to give eventually rise to a galaxy formation including ours.

In given case we had about 1500 clusters per our Galaxy, each with total mass about $10 M_{\odot}$ and contained numerous small PBHs. Initial PBH mass distribution was obtained there in the form of a power-law just like we get here $$\frac{dN}{dM_{\rm in}}\propto M_{\rm in}^{-2}. $$
It transforms to 
\begin{equation}
\frac{dN}{dM}\propto \frac{(M/M_*)^2}{\left(1+(M/M_*)^3\right)^{4/3}}
\label{dNdM}
\end{equation}
due to Hawking evaporation with $M_*\approx 5\times 10^{14}$ g being the mass of PBH evaporated by now. 
Energy spectrum of photons from PBH $I(E)=d\dot N_{\gamma}/dE$ can be approximately obtained by convolution of distribution \eqref{dNdM} with intensity for single PBH (taken as Planck black body spectrum)
\begin{equation}
\frac{d\dot N_{\gamma}}{dE}(M)=\frac{\sigma}{2\pi^2}\frac{E^2}{\exp\left(E/T\right)-1}.
\label{dNgdM}
\end{equation}

\begin{figure}[t]
    \centering
    \includegraphics[width=0.7\textwidth]{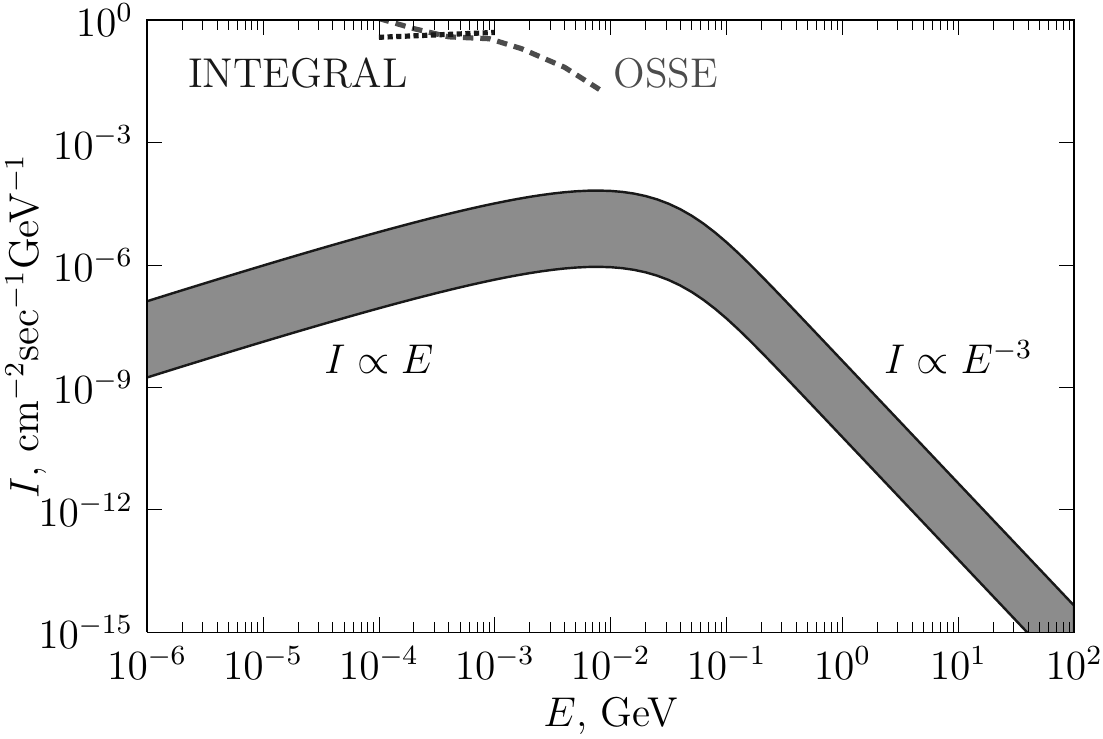}
    \caption{The expected gamma-ray spectra from PBH clusters (the shaded region).  Below the shaded region the flux from PBH is lower than 
    LAT integral sensitivity. Above it, too few clusters are expected to be observed as PGRS, namely less than one 
    (see the text for more details). The differential sensitivity of detectors INTEGRAL and OSSE is also shown.}
    \label{gamma}
\end{figure}

A total amount of photons from the cluster with $E>100$ MeV is obtained to be
$$\dot N_{\gamma}=\int\int \frac{dN}{dM}dM\frac{d\dot N_{\gamma}}{dE}dE\approx 7\times 10^{36} \text{ s}^{-1}.$$ 
Then condition
$$\frac{\dot N_{\gamma}}{4\pi r_{\max}^2}>F_{\min}$$
defines the maximum distance at which the cluster can be seen, where $F_{\min}$ is the respective Fermi lower threshold. It allows us estimating roughly possible number of PBHs which could be observed as gamma-ray sources, 
$$N_{\rm PBH}\sim n_{\rm PBH}\frac{4}{3}\pi r_{\max}^3\sim 30,$$ 
where $n_{\rm PBH}$ is the PBH cluster number density in Galaxy which was supposed to be homogeneous. So, PBH cluster at $r_{\max}$ has a minimal flux $I$ accessible to observation, which defines the lower boundary of the gray band in the Fig.\ref{gamma}. The closer the PBH clusters, the brighter they are, but the smaller their number is. At some distance only one cluster can be expected to be observed, the intensity, in this case, corresponds to the upper boundary of the band in Fig.\ref{gamma}.

In the future, one has to take into account the evaporation of light PBHs from a cluster which gives the main effect in gamma-radiation. Nonetheless, at the chosen parameters, it was found out that Fermi/LAT could observe the clusters as PGRS with power spectra as Fig.\eqref{gamma} shows. So (at least) part of the unidentified PGRS observed by Fermi/LAT could be explained by the PBH clusters. 

\begin{figure}[t]
    \centering
    \includegraphics[width=0.7\textwidth]{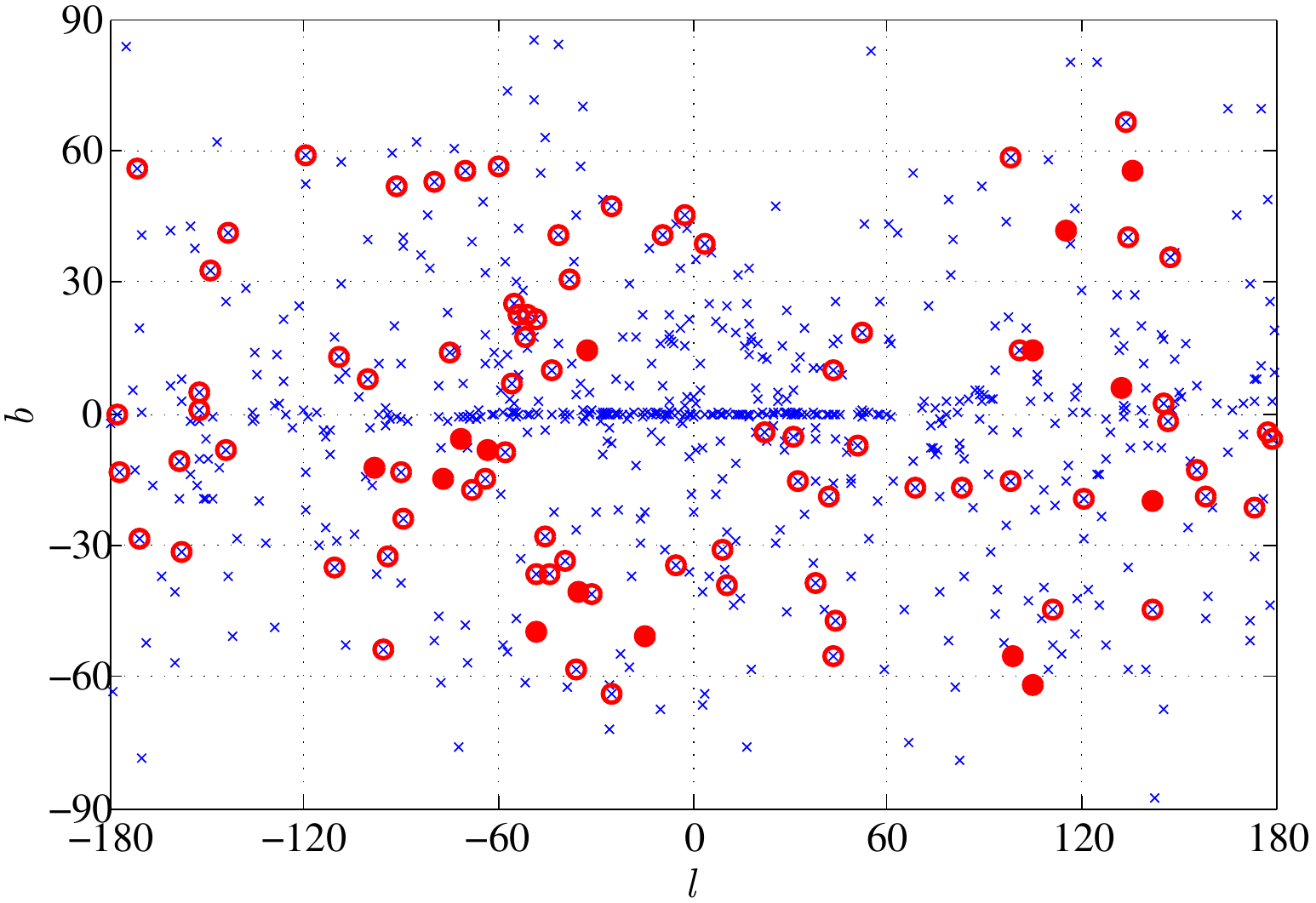}
    \caption{Unidentified PGRS seen by Fermi LAT (blue crosses) on celestial sphere (in galactic coordinates $b$ and $l$) are shown.   The red filled markers are the sources with spectrum indices 3 within $1\sigma$ error, unfilled markers are the sources with spectrum indices 3 within $3\sigma$ error.}
    \label{map}
\end{figure}

In figure  \ref{gamma}, sensitivities of X-ray telescopes (INTEGRAL \cite{2003A&A...411L...1W} and OSSE \cite{1992NASCP3137....3C}) are also shown, which turned out to be well above the expected flux which Fermi/LAT can see.

As one can see in figure \ref{gamma}, at the given parameters, the gamma-ray spectrum behaves from energy as $E^{-3}$ at high energetic part (at low one it goes as $\propto E$ because of change of PBH mass spectrum \eqref{dNdM} due to Hawking evaporation). Unfortunately, not for all PGRS observed a spectrum was measured, but one can select those of them where it was done, and select from them those with a similar spectrum, figure \ref{map}. One can note that such sources are distributed over the sky more or less homogeneously as it would be expected for clusters.

\subsection{Reionization}
\label{reion}

Reionization of the Universe happened at $z\sim 7\div 8$ is still possibly one of the problems of cosmology and astrophysics. This fact was established by different observations, mainly through Gunn-Peterson effect and CMB data (see, e.g., \cite{2016A&A...596A.108P,2001AJ....122.2850B}). There are attempts to explain it with dwarf galaxies, quasars, and stars of Population III 
\cite{2001PhR...349..125B, 2006ApJ...639..621A, 2012ApJ...752L...5B}.

PBHs could be at least one of the contributors into reionization of the Universe, and both light and massive PBHs could do it. The massive ones~--- due to radiation induced by accretion, \cite{2017PhRvD..96h3524P,2017PhRvD..95d3534A}.
However, space correlations of the X-ray and Infrared backgrounds could instead favour this opportunity \cite{2018PDU....22..137C,2016ApJ...823L..25K,2017ApJ...847L..11C}.

Light PBHs could also contribute, they can be divided into two components. As was shown above, the PBH cluster loses predominantly light PBHs during its evolution, which then may be supposed to spread more or less homogeneously in space. However, at the same time the total cluster mass changes insignificantly, so the lightest PBHs escaped the cluster make up small DM fraction and may avoid constraints coming from gamma-radiation as well as those inside it.

\begin{figure}[t]
    \centering
    \includegraphics[width=0.85\textwidth]{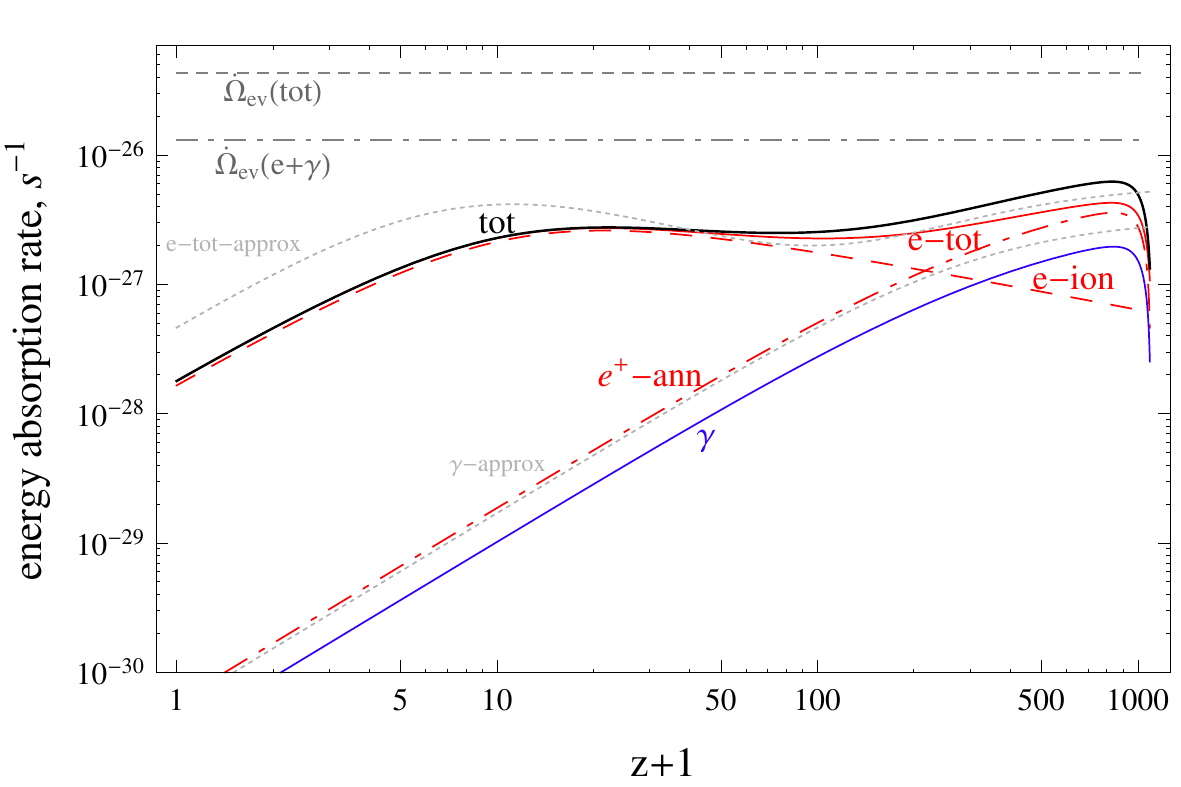}
    \caption{ Energy absorption rates for all processes considered: $\dot{\Omega}_{\rm abs}^{(e{\rm-ion})}$, $\dot{\Omega}_{\rm abs}^{(e{\rm-ann})}$ and their sum, $\dot{\Omega}_{\rm abs}^{(\gamma)}$ and the total rate, for $M=5\times 10^{16}$ g. Light gray lines relate to the respective losses obtained approximately (see \cite{2014MPLA...2940005B,2015JCAP...01..041B}). $\dot{\Omega}_{\rm ev}$ is shown to illustrate total evaporation rate into all species and $e^{\pm}+\gamma$ only.}
    \label{rates}
\end{figure}

All the light PBHs, both inside and outside the cluster, can contribute to reionization of the Universe due to Hawking's radiation. Reionization effect due to homogeneously distributed PBHs have been considered by our group \cite{ClustPBH2015,2015JCAP...01..041B,IJMPD..2015..24..1545005,2017IJMPD..2650102B}. PBH of the mass $M\sim 10^{15}\div 10^{17}$ g would produce ionizing Hawking's radiation in the form of (sub)relativistic electrons/positrons and photons (along with non-ionizing neutrinos and gravitons). They lose their energy due to different processes: $e^{\pm}$ -- due to inverse Compton scattering off CMB (Cosmic Microwave Background) photons, ionization losses, redshift; $\gamma$ -- due to Compton scattering off electrons of the medium and redshift, and photons from $e^+$-annihilation have the same losses. All the energy loss rates are shown in Fig.\ \ref{rates} depending on the redshift for monochromatic PBH mass distribution with $M=5\times 10^{16}$ g and their density as big as possible from data on the gamma-ray background.
In Fig.~\ref{rates} the value 
$$
    \dot\Omega^{(i)}_{\text{abs}}\equiv \frac{\dot\epsilon^{(i)}_\text{abs}}{\rho_{\rm crit}}
$$ 
is referred to absorption rate, where $\dot\epsilon^{(i)}_\text{abs}$ is the rate of energy deposition in baryonic matter per unit volume for $i$-th process, $\rho_{\rm crit}$ is the critical density of the Universe.

As one can see from the figure, these are $e^{\pm}$-ionization losses which give the main contribution to energy transfer from PBH radiation to the matter.
Re-ionization effect is reached in given approximation for PBH mass around $(3..7)\times 10^{16}$ g. 

The effect can be enhanced if one uses less trivial mass distributions, e.g., 
two-peak or power-law ones \cite{2017IJMPD..2650102B}. Reionization (enhancement of contribution in it) could be reached at a quite noticeable region of parameters for a power-law mass distribution (see Fig.~3 left of the reference above).

However, within our approximation, it was supposed that all energy deposits in matter via a single channel -- heat, while it should be shared between three -- heat, ionization (what goes to overcoming of ionization potential), atom excitation \cite{2017JCAP...03..043P,2017PhRvD..95h3006C,Liu:2016cnk,Slatyer:2015kla,HyeRec}. 
The effect could be weaker if we took into account all the channels.

Also one should take into account the data on baryon temperature at $z\sim 17$, deduced from 21 cm absorption line of hydrogen observation \cite{Nat..555..67..2018}, which may give extra restriction for PBH in a given mass range \cite{2018arXiv180309390C}.

\section{Conclusion}
The necessary components of the modern theory of the structure and evolution of the Universe are the inflation, the baryosynthesis, and the dark matter/energy.  The physical basis of this currently Standard cosmological paradigm implies physics beyond the Standard Model of fundamental natural forces. The extension of the Standard Model of elementary particles is necessary to explain the mechanisms of inflation and baryosynthesis, as well as physics of dark matter and dark energy, inevitably contains additional cosmologically viable consequences. Primordial black holes are among the most important probes for such features of new physics in the very early Universe.

The progress in observations of distant object unambiguously points to the early formation of the massive and supermassive black holes. The origin of black holes of 10-30 Solar masses, gravitational wave signal of whose coalescence was detected by advanced LIGO and Virgo, or the origin of massive black hole seeds for AGNs may cause problems for their explanation based on stellar evolution in galaxies and can find their nature in primordial black holes.

A substantial amount of models are elaborated to explain the phenomena of primordial black hole formation. Some of them contain an opportunity of PBH cluster formation. The set of such models is discussed in this review. We focus mainly on one of them related to the phase transition with the formation of the closed walls. There are three steps of the PBH formation.  Firstly, the quantum fluctuations during the inflation produce the space areas with scalar field values at both sides of a potential extremum. Secondly, after the end of inflation, the field settled in different vacua forming the closed walls. At the third stage, these walls move through the surrounding media, heating it and finally forming the black holes. 

Field fluctuations at successive steps of inflation lead to the change of vacuum at smaller scales so that after the second phase transition the largest wall becomes inevitably accompanied by a cloud of smaller walls. Provided that the width of the wall doesn't exceed its gravitational radius, the walls collapse into smaller black holes, so that PBH cluster is formed.

Potential properties and initial conditions determine the range of PBH masses in clusters. The minimal mass is determined by the condition that the gravitational radius of the wall exceeds its width, while the condition that the wall doesn't start to dominate before it enters the horizon, puts an upper limit on the PBH mass.

Inevitable ingredient of the model is the internal evolution of the PBH cluster. After a cluster separates from the expansion, the formation of BH binaries inside it is strongly facilitated as compared with the case of stochastic PBH distribution. The PBHs merge and scatter each other, some of them evaporate from the cluster. The first numerical simulations presented in this review indicate that space and the mass distributions are not altered significantly. We show in the section~\ref{InternalDynam} that only $\sim15\%$ of black holes leave the cluster. Remaining BHs form the compact cluster with sizes $\sim1$~pc and the total mass $\gtrsim 10^3 M_\odot$ at the moment $z\sim20-40$.  Such clusters could serve as the <<seeds>> of the early quasars observed at $z>6$.

The features of PBH clustering can shed new light on the observational data and provide us with new degrees of freedom in their analysis.
PBH cluster model inherits all the advantages of those models predicting uniform PBH distribution like a possible explanation of the existence of supermassive black holes (origin of the early quasars), binary BH merges registered by LIGO/Virgo through gravitational waves (GW), contribution to reionization of the Universe, but also has additional benefits. The cluster could alleviate or avoid at all the existing constraints on the uniformly distributed monochromatic PBH abundance so possibly becoming a real DM candidate. Most of the existing constraints on PBH density should be re-considered if to apply to the clusters. The first attempts, mentioned in this review, have been already done to limit possible abundance due to observed BH merger rate,
but PBHs of respective mass range could be much below that limit. Also unidentified cosmic gamma-ray point-like sources could be (partially) accounted for by them. One can conclude that it seems to be a much more viable model as compared with the models predicting uniform PBH distribution.

It is worth noting that the gravitational waves
can become the specific prospective tool for the cluster model validation. The PBHs merges into pairs into the clusters could provide the sources of GW bursts in addition to the binary black holes of stellar origin. One of the most interesting effects is the possibility of recurrent (multiple) events of GW bursts from the same sufficiently dense cluster. On the contrary, the binary systems of stellar origin produce only single events from different galaxies. Therefore, future observations of the anticipated multiple events can provide the decisive test of the PBHs clustering discussed in this review.

The discussion made above is applicable to any compact cluster model. To date, several such models, different from the one we basically focused on here, have been proposed.
Their formation mechanisms are based on strong primordial density fluctuations. According to these models, PBHs are clustered in the course of evolution due to their Poisson distribution or due to the seed effect. Such mechanisms are also applied to big clusters on the scale of large scale structure to explain its formation.

The mechanism of the PBH cluster formation discussed in this review can be tested by future observations. Indeed, the closed walls are its necessary attribute. They are nucleated having the shape far from the spherical one \cite{PBH_2,2002astro.ph..2505K, 2006PhRvD..74f3504B, 2017PhRvL.119s1103B}. The wall inhomogeneities are damped transferring the energy to the surrounding media and heating it. According to our estimation, the heating could be extremely effective so that the temperature of matter around such PBH cluster could be high. This allows one to separate our mechanism from the others. We develop the theme, understanding its importance.

In the context of cosmoparticle physics, studying the fundamental relationship of cosmology and particle physics, search for PBH clusters and their study acquire the meaning of nontrivial probe of new physics of the very early Universe.

\section{Acknowledgement}

K.M.B.\ would like to express his special gratitude to Maxim Laletin for providing 
many useful references and fruitful discussions. The work of MEPhI group on studies of the formation of PBH clusters was supported by the Ministry of Education and Science of the Russian Federation, MEPhI Academic Excellence Project (contract № 02.a03.21.0005, 27.08.2013). The work of K.M.B. on observation effects of PBHs is also funded by the Ministry of Education and Science of the Russia, Project № 3.6760.2017/BY. The work of V.I.D. and Yu.N.E. on the evolution of PBH clusters was supported by the Russian Foundation for Basic work Research Grant
№ 18-52-15001. The work of A.A.K., L.A.K., V.V.N. and E.A.E. on the $N$-body simulation is funded by the Russian Foundation, Project № 19-02-00930.
The work of S.G.R. on initial conditions of early Universe was also supported by the Ministry of Education and Science of the Russian Federation, Project № 3.4970.2017/BY and by the Russian Government Program of Competitive Growth of Kazan Federal University. The work by M.Yu.K. on indirect effects of dark matter physics was supported by grant of Russian Science Foundation (project №~18-12-00213).

\bibliographystyle{JHEP}
\bibliography{bib2.bib}
\end{document}

%% file: Formation/Formation.tex
\section{PBH formation}
\label{sec:PBH_form}

One of the reasons for primordial black holes formation is phase transitions during and after inflation. This is three-step process. As discussed in the Introduction, closed walls connecting two vacuum states appear after the inflation is terminated. The shape of most of them is far from beeing spherical.  At the second step the walls evolve in the plasma of relativistic particles that are produced by the reheating stage of the inflation. Internal tension leads to decreasing the surface square transmitting its potential energy into the kinetic energy of the wall. In its turn, the kinetic energy is also decreasing due to friction of the surrounding matter. As the result, the wall minimizes its surface energy by acquiring spherical shape and shrinking. This step is highly model dependent. To proceed, we suppose that the wall acquires the spherical shape soon after it crosses the horizon. The third step consists of wall shrinking due to its internal tension. It is finished by a black hole creation if the gravitational radius of the wall is larger than the wall's width.

One can conclude that production of closed walls lead to strong inhomogeneities able to collapse into PBH. Multiple quantum fluctuations of a scalar field result in PBHs clusters formation, with a scalar field not necessarily being an inflaton. In the review we will focus basically on this scenario of PBHs formation discussed in \cite{PBH_1,PBH_2,Mechanism_BH}. 

The mechanism of the PBH cluster formation is represented in subsections \ref{sec:2.1.Mult_fluct}, \ref{sec:2.2.Distrib_reg}, \ref{sec:2.3.Distrib_reg} in general form so that it can be used in a wide class of potentials. In subsection \ref{specific model} we perform numerical analysis on the basis of specific model. This model is also used in Section \ref{sec:PBH_cluster_form}. It is shown there that the mass spectrum of cluster can be varied in wide range depending on Lagrangian parameters. Starting from Section \ref{sec:Intern_dynam} we discuss different applications for the PBH clusters characterized by different mass spectrum.

\subsection{Multiple fluctuations on inflationary stage}
\label{sec:2.1.Mult_fluct}

Consider a scalar field $\phi$ with potential possessing at least one maximum or a saddle point. As a result of inflation the observable Universe is formed from a single causally connected region with a characteristic size of $H^{-1}$. Let our Universe be nucleated with an initial field value $\phi_\text{u}$ close to a potential maximum or saddle point.  Due to the classical motion, most of the space will be finally filled with a field lying in a specific minimum. Nevertheless, quantum fluctuations during the inflation gradually bring the field over the potential barrier in individual space regions. After inflation, classical field evolution in such space regions rolls the field down to another minimum, resulting in the formation of a wall between such domain and the outer Universe.
For future discussion, let us denote as $\Phi$ a set of those field values $\phi_c$, (i.e. $\phi_c \in \Phi$) that lead to the formation of the space islands (domains) after the inflation is finished.

Multiple quantum fluctuations during inflation can be described as random walks \cite{Quant_to_Class}. One can solve the Fokker-Planck equation for the field $\phi$ \cite{FP_Solve2} and get $f(\phi, t)$~--- field $\phi$ values distribution by the moment of time $t$. Hereinafter we assume by definition the relation $N\equiv Ht$ of the number of e-folds $N$ to the time $t$ passed from the beginning of inflation. Neglecting the shape of the potential (for more detailed calculations see \cite{FP_Solve1,FP_Solve2,FP_Solve3,Mechanism_BH,2017JCAP...10..018H}) and considering $\phi_\text{u}$ as an initial field value at the beginning of inflation one can get the probability density of finding field $\phi$ at an arbitrary space point
\begin{equation}\label{f_ot_t}
f(\phi,t)=\frac{1}{\sqrt{2\pi}\sigma(t)}\exp\left(-\frac{(\phi-\phi_\text{u})^2}{2\sigma^2(t)}\right), \quad \sigma(t)=\frac{H}{2\pi}\sqrt{Ht}.
\end{equation}
The formula \eqref{f_ot_t} is valid for the field $\phi$, a mass of which can be neglected in comparison with the Hubble parameter during inflation, \cite{linder1990particle}. The corresponding formula for $m \neq 0$ was obtained in \cite{Primord_Inhomo,2007arXiv0709.0070D}.

The probability of finding the field within the region $\Phi$ by the moment of time $t$ is:
\begin{equation}\label{Pt}
P(t)=\int\limits_{\Phi} f(\phi,t)\, d\phi = 
\int\limits_{\phi_\text{cr}}^{\infty} f(\phi,t)\, d\phi = 
\frac{1}{2}\erfc\left(\frac{\phi_\text{cr}-\phi_\text{u}}{\sqrt{2}\sigma(t)}\right),
\end{equation}
where $\phi_\text{cr}$ is the boundary field value, separating the region $\Phi$ in a case of one-field potential.

The space of the size of the Hubble radius is divided into $n_\text{u}(t)=e^{3Ht}$ causally independent patches by the moment $t$. Hence, the number of critical fluctuations (or the number of critical regions) is: 
\begin{equation}\label{crit}
n_{c}(t)=P(t) e^{3Ht}.
\end{equation}
The size of fluctuations is limited by the size of the horizon $H^{-1}$. Hereinafter we assume the number of critical regions to be much less than the number of all causally independent regions. Otherwise walls abundance would contradict the observable rate of the Universe expansion and the CMB data.

\subsection{Distribution of regions over the size}
\label{sec:2.2.Distrib_reg}

Let the critical fluctuation takes place at the moment of time $t$ from the beginning of inflation. After inflation, its size will be:
  \begin{equation}\label{rinfott}
r_\text{inf}(t)=H^{-1}e^{N_\text{inf}-Ht},
\end{equation}
where $N_\text{inf}\simeq 60$ is the total number of e-folds needed for the formation of the visible Universe. After the inflation is finished, the field tends to different minima in different areas. Main part of the Universe contains the field in a specific minimum while the regions filled by the critical fluctuations form the domains of another vacuum separated from the surrounding space by walls. As has been shown in \cite{2017JCAP...12..044D, 2017JCAP...04..050D}, 
the walls are not spherical just after formation. On the contrary, the fractal-like surface is much more natural for this mechanism. Here we suppose that the energy of the wall fluctuations is quickly transferred to matter that leads to spherical walls surrounded by heated matter.

Closed wall starts to collapse after crossing its horizon which is expanding on the RD-stage as $r_\text{hor}(\tau)=2\tau$, where $\tau$ is the period of time passed after the beginning of the RD-stage. The domain wall itself expands as $r(\tau)=r_\text{inf}\sqrt{\tau/t_\text{inf}}$ on this stage, where $t_\text{inf}=N_\text{inf}/H$. By defining the moment of time $\tau$ when the horizon reaches the size of the wall, and substituting it into $r(\tau)$ we get the size of the wall at the beginning of the collapse:
\begin{equation}\label{rott}
r(t)\simeq\frac{r_\text{inf}^2}{2t_\text{inf}}\simeq\frac{e^{2(N_\text{inf}-Ht)}}{2H N_\text{inf}}.
\end{equation}

The expression \eqref{rott} relates the moment of the domain formation during inflation $t$ to the final size of the domain $r$. By substituting $t\equiv t(r)$ into \eqref{crit} we get the distribution of walls over the size
\begin{equation}\label{notr}
n_{c}(r)=P\big(t(r)\big) e^{3H t(r)}.
\end{equation}

\subsection{Distribution of regions over the mass}
\label{sec:2.3.Distrib_reg}

The fluctuating mechanism discussed here leads to the PBH formation caused by the closed walls collapse. The PBHs mass spectrum is the result of complex processes that take place during inflation, just after the inflation and is strongly model dependent. Indeed, walls are created having the form far from the spherical one \cite{PBH_2,2002astro.ph..2505K, 2006PhRvD..74f3504B, 2017PhRvL.119s1103B}. Their small inhomogeneities are damped transferring the energy to the surrounding media and heating it. The larger perturbations of wall surface survive by the moment $t_h$ of the horizon crossing. The nonspherical collapse with the BHs in final stage has been discussed in \cite{2006PhRvD..74f3504B, 2017PhRvL.119s1103B}. 

For the spherical walls, there are three regimes of walls evolution depending on their surface energy density $\mu$. If the wall energy is too small, it cannot be concentrated to form the black hole (see formula \eqref{botlim} and discussion around it). If the wall is too massive (<<supercritical>> in terms of \cite{2017JCAP...04..050D}), inequality 
\begin{equation}\label{supcrit}
t_h > t_{\mu}\equiv \frac{M^2_{Pl}}{2\pi\mu}
\end{equation}
holds and according to \cite{2017JCAP...04..050D}, such a wall leads to the creation of baby universes with wormholes inside them. For  an  exterior  observer, a  black  hole  is  formed. Here $t_h$ is the time scale when the horizon reaches a wall size.

A wall for which $t_h < t_{\mu}$ is called “subcritical”. Its gravitational field can be  neglected before horizon crossing the wall scale. If its form is near spherical, it shrinks to a size smaller than the corresponding Schwarzschild radius, forming a black hole with ordinary internal structure. The final masses of the black holes depending on the initial data and parameters of the model are also analyzed in  \cite{2016JCAP...02..064G, 2017JCAP...12..044D, 2017JCAP...04..050D}.

For the qualitative analysis, we suggest the case of the subcritical spherical walls. Also, we consider the case of the field potential with the energy-degenerated minima.

Radius of a spherical wall with the mass $m$ is $r(m) = \sqrt{m / 4\pi \mu}$, where $\mu$ is the surface energy density. The latter depends on specific form of the potential. By substituting $r\equiv r(m)$ into \eqref{notr} we get the distribution over the mass:
\begin{multline}\label{tvr}
n_c(m) = P\Big(t\big(r(m)\big)\Big) e^{3H t\big(r(m)\big)}= \\
= \frac{1}{2}\exp\left({\frac{3}{2}\left[2N_\text{inf}-\ln{\left(HN_\text{inf} \sqrt{\frac{m}{\pi\mu}}\right)}\right]}\right)\times\\ 
\times\erfc\left(\cfrac{2\pi (\phi_\text{cr}-\phi_\text{u})}{H\sqrt{2N_\text{inf}-\ln{\left(HN_\text{inf} \sqrt{\dfrac{m}{\pi\mu}}\right)}}}\right).
\end{multline}

The expression \eqref{tvr} is the direct consequence of expression \eqref{crit}. 
It gives the total number of critical regions that have arisen by the time $t$. Therefore expression \eqref{tvr} denotes the total number of PBHs with masses greater than $m$~--- the integral distribution. Differential distribution can be obtained by differentiating expression \eqref{tvr}

\subsection{Model with specific shape of the potential}
\label{specific model}

Let us illustrate the speculations above by a specific potential satisfying the necessary conditions - the presence of extremum. Consider a complex scalar field with the potential
\begin{equation}
	\label{Mexica}
	V(\phi)=\lambda\left(|\phi|^2-f^2/2\right)^2 + \Lambda^4(1-\cos \theta),
\end{equation}
where $\phi=re^{i\theta}$. This field coexists with an inflaton field which drives the Hubble constant H during the inflationary stage. Last term in \eqref{Mexica} reflects the  renormalization effects to the Lagrangian (see details and refs in \cite{1993PhRvD..47..426A}) and has been used in the model of the baryon asymmetry observed in the Universe \cite{1997PhRvD..56.6155D}. This term is assumed to be negligible at the inflationary stage.

The parameter $\Lambda$ appears as a result of the instanton effects and renormalization, so that its value cannot be large and we assume that $\Lambda\ll H,f$. The second term in \eqref{Mexica}  plays a significant role in the post-inflationary stage, when the Hubble parameter decreases with time (e.g. $H = 1/2t$ during the radiation dominated stage, $H= 2/3t$ during the matter dominated stage).

When the inflation is finished, the field is captured in the potential minima $\phi=\frac{f}{\sqrt{2}}e^{2\pi n i},\quad n=0,1,2,\ldots$. If neighbour space areas are filled by different vacuum states, say $\theta=0$ and $\theta =2\pi$, the second term in \eqref{Mexica} leads to the effective Lagrangian
\begin{equation}
 \label{Lchi}
 L_{\chi}=\frac12(\partial \chi)^2 - \Lambda^4 (1-\cos(\chi/f))
\end{equation}
for the dynamical value $\chi = f\theta$.
The one-dimensional kink solution for the cosine potential is well known \cite{Rajaraman1982}: 
\begin{equation}
 \label{kink}
 \theta(z)=-4\arctan\left[e^{\frac{\Lambda^2}{f}(z-z_0)}\right].
\end{equation}
It is assumed throughout the paper that the wall width $d$ is much smaller than the wall size. Therefore the wall is almost plane for an observer near the wall and we can apply formula \eqref{kink} for describing the distribution of phase across a plane border separating two areas with different vacuum states. Knowledge of solution \eqref{kink} allows one to find the wall's surface energy density 
\begin{equation}
 \label{mu}
 \mu=4\Lambda^2 f.
\end{equation}
Energy of this field configuration is concentrated in the plane of width $d = 2f/\Lambda^2$ forming a wall with the center at $z_0$.

The domain size immediately after the end of inflation
markedly exceeds the horizon size at the FRW expansion stage. The overall
contraction of the closed wall may begin only when the horizon size will be
equal to the domain size $R_{w}$ with total energy $E_w\sim 4\pi R_w^2\mu$. Up to this moment, the characteristic domain size increases with the expanding Universe. Evidently, internal stresses developed in the wall after crossing the horizon and external friction leads to minimization of the wall's surface. This implies that, having entered the horizon, the wall tends, firstly, to acquire a spherical shape and, secondly, to contract toward the center. The wall contracts up to the minimal size of the order of the wall width $d$. A black hole can be formed if this scale is smaller than the gravitational radius of a wall. Following \cite{Khlop_Rub_2004}, 
the black holes are formed under condition
\begin{equation}
 \label{botlim}
 d\lesssim r_g\simeq 2E_w/M_{Pl}^2.
\end{equation}
Here the Planck mass is denoted as $M_{Pl}$. Indeed, if the width of the wall exceeds its gravitational radius, the energy of the wall can not be concentrated within it. The wall size $d$ depends only on the physical parameters of the Lagrangian while the total wall energy can be macroscopically large. The inequality \eqref{botlim} gives lower limit for the black hole mass.

We choose the following values of the model parameters, which agree with the data on the observable CMB anisotropy: $H=10^{13}$~GeV, $N_\text{inf}=60$, field parameter $\Lambda=0.05$GeV, and initial value $\theta_\text{u}=0.05 \pi$. The distribution is very sensitive to the parameter values. 

\begin{figure}[t]
 \centering
 \includegraphics[width=0.8\linewidth]{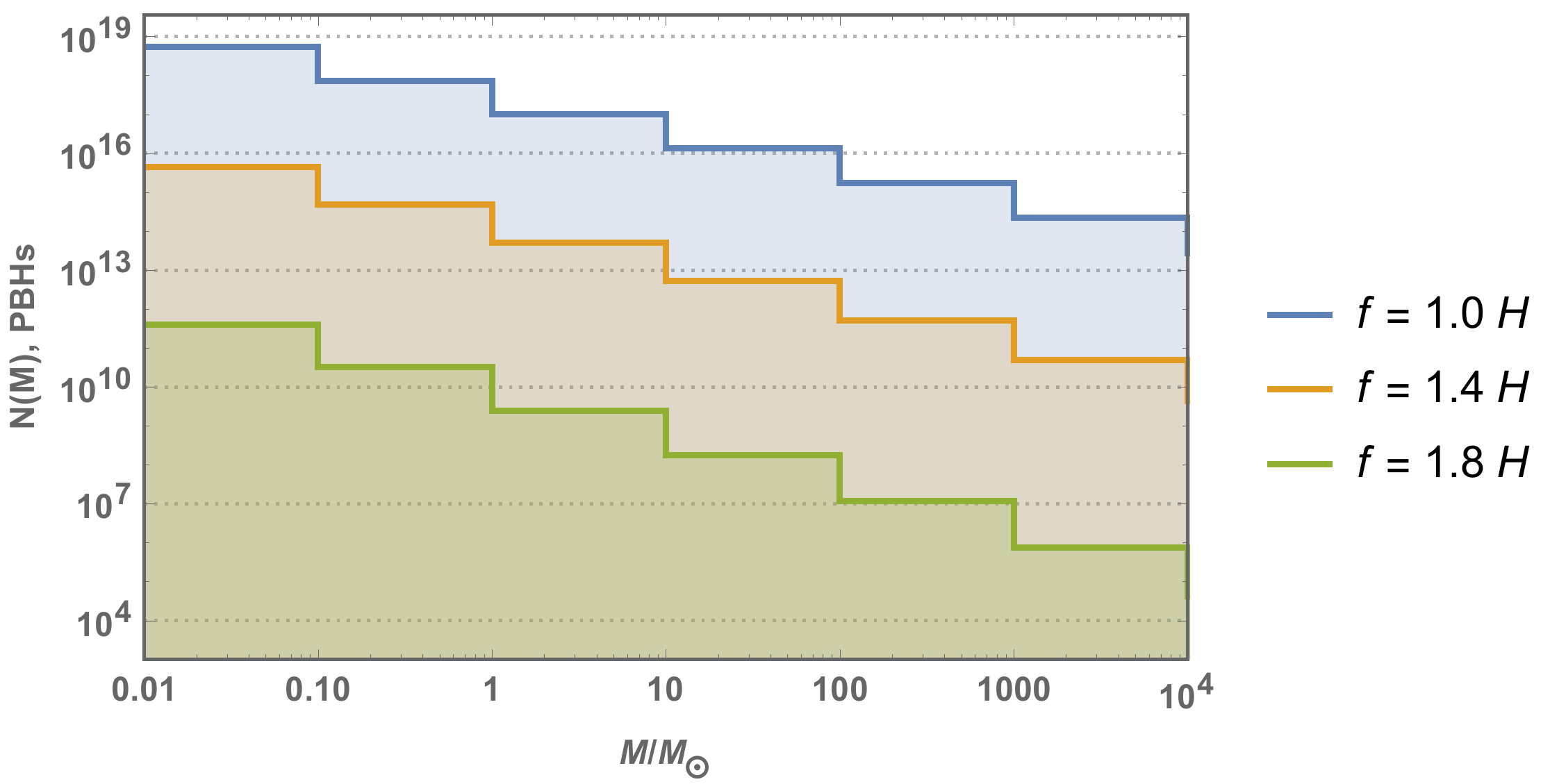}

 \caption{PBHs mass spectrum for the visible Universe: $f=1.4 H$ line represents realistic PBH distribution. In this case, the number of PBHs with a mass $\sim 10^2 M_{\odot}$ corresponds to the number of galaxies in the observable Universe. The role of other lines is to demonstrate a strong dependence of the distribution on the parameter $f$.}

 \label{fig:NMUniv}
\end{figure}

It can be seen (Fig.~\ref{fig:NMUniv}) that number of PBH depends crucially on the vacuum expectation value $f$. Moreover, the number of PBH is affected significantly by the initial field value $\theta_U$: the closer is the initial value to the critical value $\pi$~--- the more PBH of all masses will be in the Universe. The wall's density, which depends on a value of the parameter $\Lambda$, also affects how massive PBH can be formed, but this affection is not as significant as the changing in $\theta_U$ or $f$.

Domains with low masses do not collapse into black holes because their gravitational radius is less than the thickness of their wall, see inequality \eqref{botlim}. With the chosen model parameters values, the border is located at the mass of $\sim 10^{-2}  M_{\odot}$ as is shown in Fig.\ref{fig:NMUniv}. The considered mechanism of PBH formation leads to nontrivial situation when massive PBH can exist along with total absence of lower masses PBH. Too large walls begin to dominate locally in the Universe at the RD-stage, which causes them to stop expanding with a subsequent collapse into the PBH as follows from \eqref{supcrit}. Therefore, there is an upper limit on the size of the walls and, as a consequence, on the mass of the PBH. With the chosen model parameters values the upper border is located at the mass of $\sim 10^{4} M_{\odot}$.

\section{PBH cluster formation. Internal structure of cluster}
\label{sec:PBH_cluster_form}

As has been shown in \cite{PBH_2} black holes are created in a company with substantial amount of smaller masses BHs. These walls form their common gravitational well where they merge to create BHs of larger masses. The subsequent evolution of such PBH cluster depends on its space/mass distribution, Lagrangian parameters and initial conditions. 

One has to conclude that the formation  of the PBH cluster is a complicated multi-step process. In this review, we will base on the assumption that a part of the closed walls are able to get rid of the surface perturbations and acquire a spherical shape.

Additional remark is necessary. The mechanism of closed wall formation is based on the quantum fluctuations of a scalar field near maximum of the potential $V(\phi)$ where the potential derivative is small. This means that the classical motion of the field is also slow. At the same time, the energy density fluctuations $\delta\rho/\rho \propto 1/|\dot{\phi}|$. The immediate conclusion is that the energy density fluctuations increase rapidly in the area of interest that could lead to another mechanism of BH formation acting simultaneously with the main one. In fact, the model \cite{1998PhRvD..58h3510Y} is based on this mechanism.

As it has been told earlier, critical region will be formed with higher probability due to multiple fluctuations instead of direct tunneling. Hence, the closer the field value inside the space region-predecessor is to the critical value, the higher is the probability of the critical region formation. Thus, the probability of new PBH formation around the existing one (the parent PBH) increases. PBHs form clusters with fractal structure. Subsequent evolution of these clusters lead to the merging of black holes into more massive ones.

\subsection{PBH mass and distance distributions inside the cluster}

Fractal structure of clusters implies that around each PBH there is an aggregate of smaller PBHs, around each of which~--- an aggregate of even smaller PBHs and etc. To study space and mass structure of a cluster let us choose some <<seed>> (parent) PBH with the mass $M_0$ and consider a cluster  of smaller PBHs around it. Let us fix a size $r_\text{cl}$ of this region by the end of inflation. Then according to \eqref{rinfott}, the moment of time when this region was formed is $t_\text{cl}=t(r_\text{cl})$. We also need the moment of time when the main PBH was formed. It can be derived from \eqref{rott}: $t_0=t(r(M_0))$. As long as such a PBH should be the only one to form, during the period of time $t_\text{0}-t_\text{cl}$ only one fluctuation should occur. One can write down the following equation by using \eqref{crit}:
\begin{equation}\label{condition}
n_c\big( \phi, t_\text{0}-t_\text{cl} \big)=1.
\end{equation}

Condition \eqref{condition} connects the size of a chosen region $r_\text{cl}$ to the initial field value $\phi$ inside it in the way that only a single PBH with the mass $M_0$ forms in this region after inflation. By solving it one can get the dependence $\phi(M_\text{0},r_\text{cl})$. Now we can calculate the spatial-mass distribution of PBHs inside specific region. It is described by the expression derived from \eqref{crit} by substituting the initial field value $\phi(M_\text{0},r_\text{cl})$ and time $t-t_\text{cl}$:
\begin{equation}\label{ncl}
n_c\big(\phi(M_\text{0},r_\text{cl}), t(r(m))-t(r_\text{cl})\big)=n_c(M_\text{0},r_\text{cl},m).
\end{equation}

Thus, by defining the mass of a <<seed>> PBH one can obtain the distribution of the rest PBHs in the region $r_\text{cl}$ over the mass. It is worth noting that similarly to \eqref{tvr} this distribution is integral, i.e. gives the number of PBHs with masses higher than $m$ inside the region $r_\text{cl}$. Considering \eqref{ncl} as a function of $r=r_\text{cl}$ one can obtain space-mass distribution.

\begin{figure}[t]
   \includegraphics[width=0.5\linewidth]{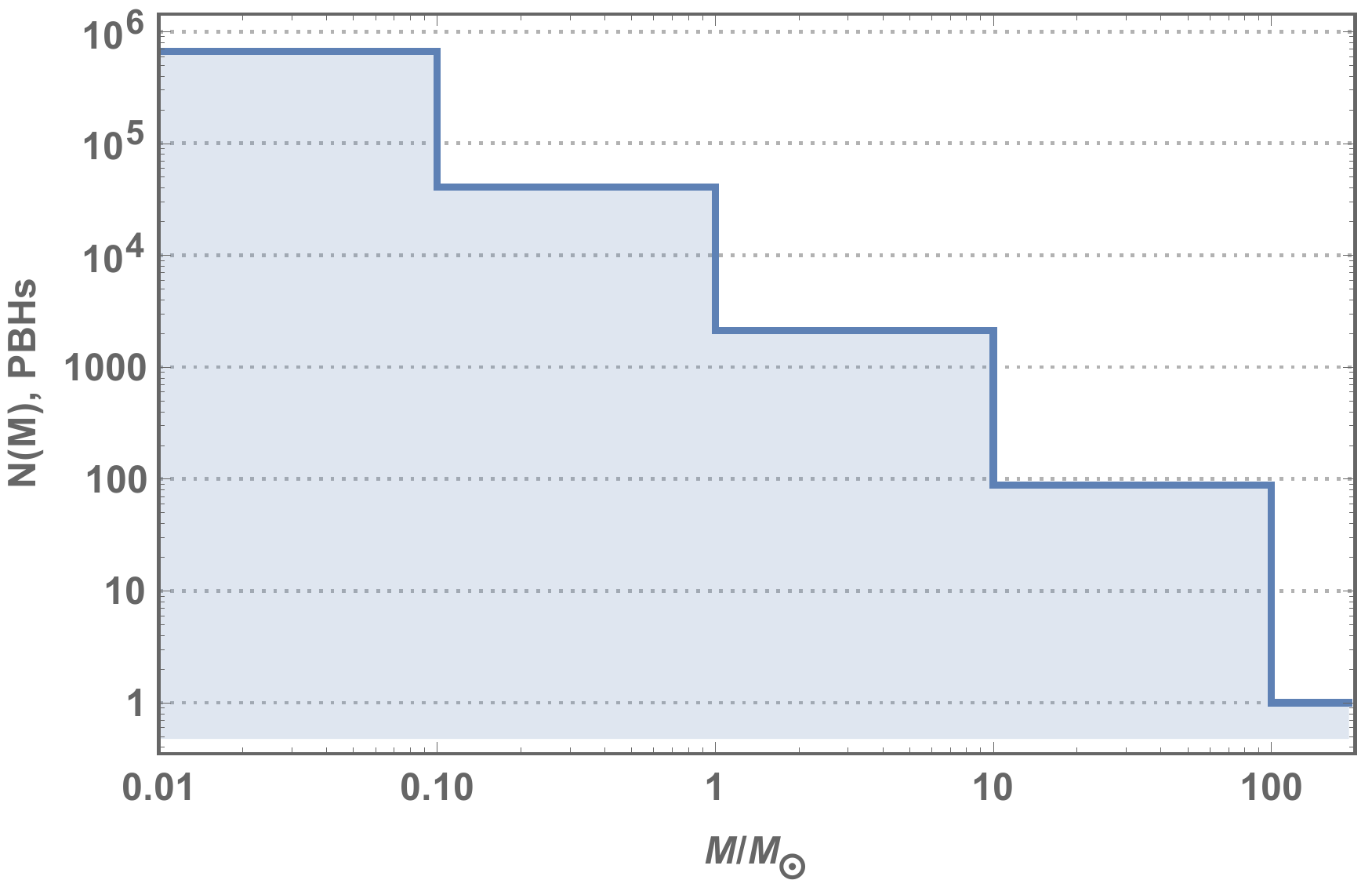}
   \caption{The PBH mass distribution for the cluster with a central PBH of the mass $10^2 M_{\odot}$. with the model parameter value $f=1.4H$. Y-axis shows the total number of PBHs in each mass interval.}
   \label{fig:Mass_Distrib}
\end{figure}

\begin{figure}[t]
 \begin{center}
  \subfigure[]{	
   \includegraphics[width=0.478\textwidth]{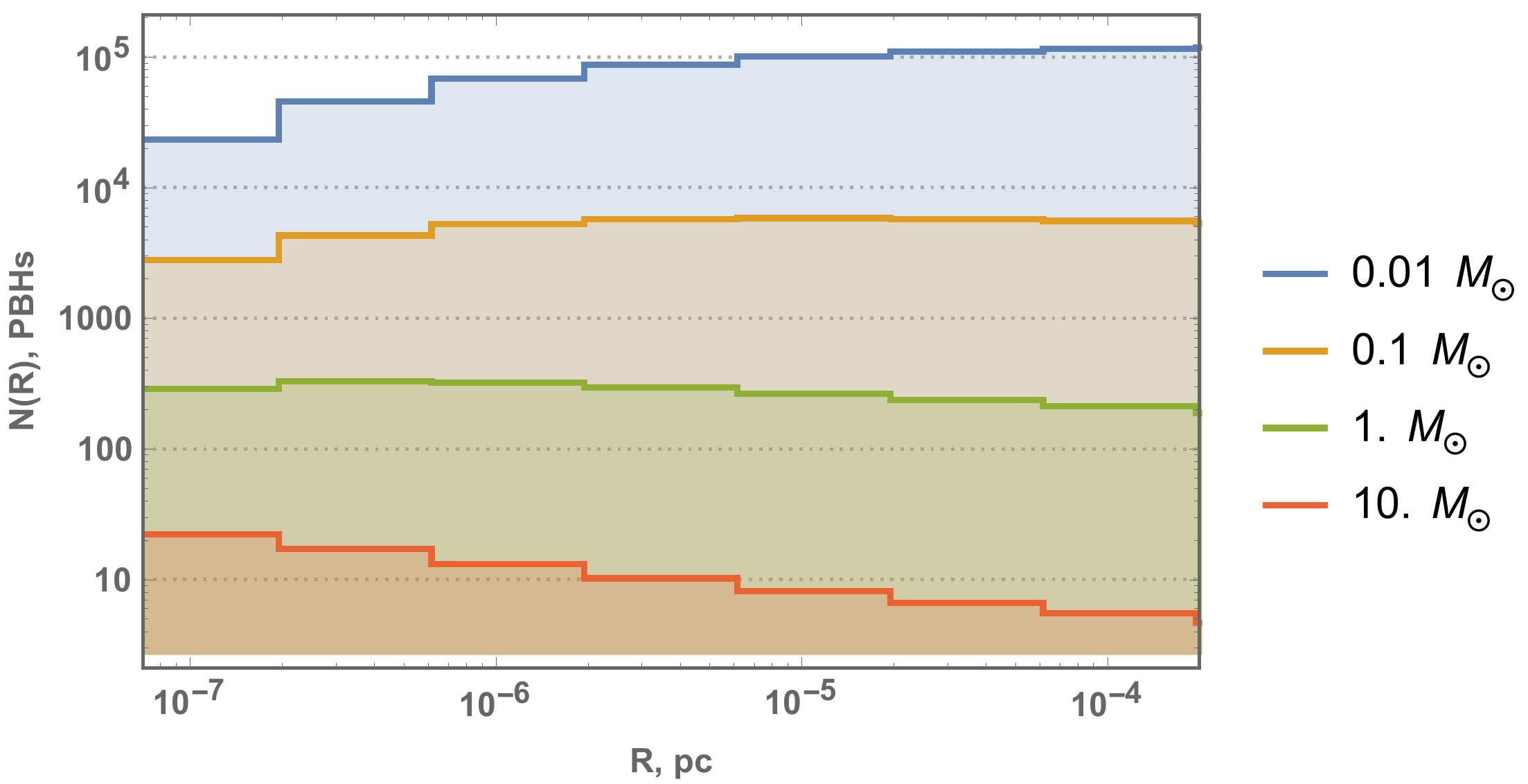}
   \label{fig:Space_Distrib}
  }
  \subfigure[]{	
   \includegraphics[width=0.478\textwidth]{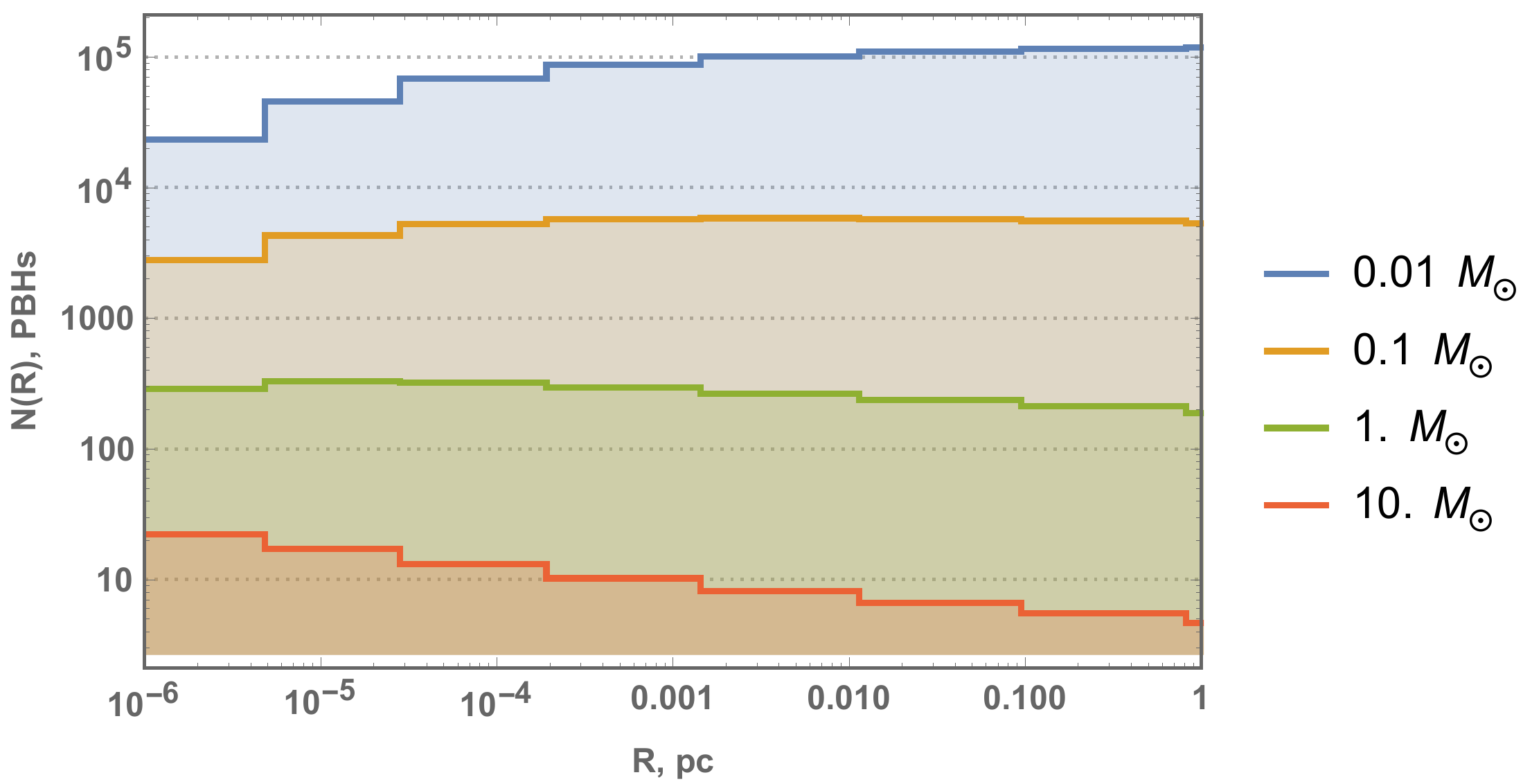}
   \label{fig:N_r_detach}
  }
 \caption{Spatial distribution inside the cluster.  The mass of central PBH is $10^2 M_{\odot}$, the selected parameters of the potential: $f=1.4H$, $\Lambda=0.05$, $\theta=0.05\pi$. (a): The distribution of PBHs inside the cluster just after its formation (after the collapse of all closed walls). (b): The distribution of PBHs just after their detachment from the Universe expansion (at $z \sim 10^4$). Characteristic cluster radius is $\sim 1$ pc. 
After detachment, the central region appears to be more dense (though it is not clearly seen in the figure) while the total number of the PBHs inside the cluster remains almost the same.}
 \label{fig:PBH_Distrib}
 \end{center}
\end{figure}

Consider as the example specific PBH cluster with a <<seed>> black hole of the mass $10^2 M_{\odot}$. The calculations were performed on the basis of model \eqref{Mexica}.
Spatial and mass PBH distributions inside the cluster are shown in Fig.~\ref{fig:Mass_Distrib},\ref{fig:PBH_Distrib}. The distributions in the Fig.~\ref{fig:Mass_Distrib}, \ref{fig:Space_Distrib} are obtained by calculating \eqref{ncl} with partitions into ranges in double-logarithmic scale. The distribution in the Fig.~\ref{fig:N_r_detach} is obtained by calculating the sequential detachment of the spherical layers of the cluster from the Universe expansion. The procedure and theory of the calculation is discussed in the next section.

%% file: Detachment/Detachment.tex
\begin{figure}[t] 
 \centering
 \includegraphics[width=0.7\textwidth]
{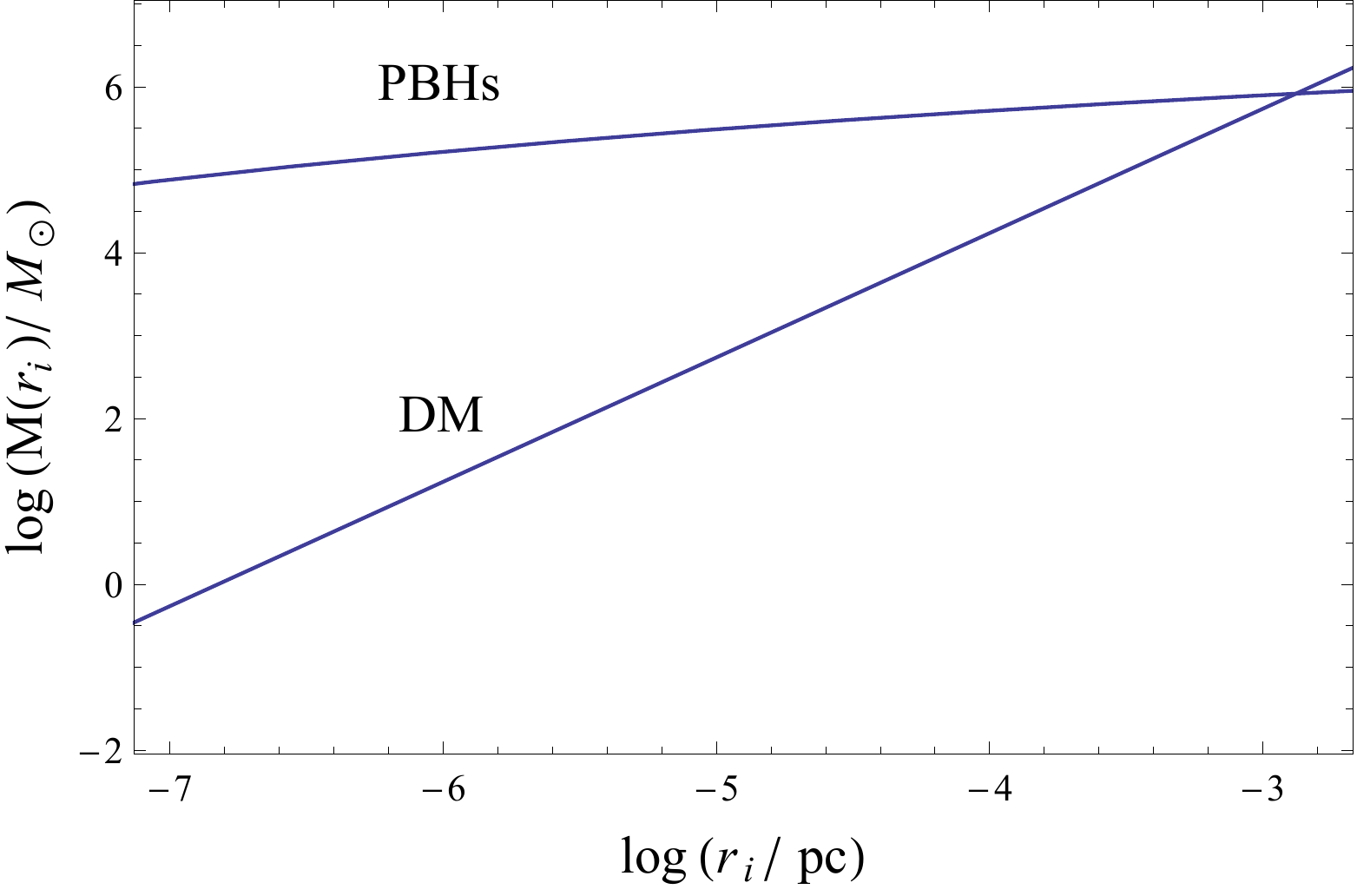}
 \caption{\label{massprof1} The initial mass profile $M_h(r_i)$ of the BHs cluster and mass profile $M_{\rm DM}(r_i)$ of dark matter halo enclosing the cluster. The radius $r_i$ is the physical radius of the shell at the moment $t_i$ of its horizon crossing, therefore the different shells are shown at different moments. That's the reason the dark matter density does not follow the homogeneous law $M_{\rm DM}\propto r^3$. The parameter values (in GeV): $\Lambda=5, f=1.6H$} 
\end{figure}

Let us consider a detachment of large cluster from the mean Hubble flow. In this case the result is more clear and affects the dark matter distribution. We describe the gravitational dynamics of some particular spherical layer, after the moment when this layer has entered under the cosmological horizon, i.e. its radius $r<ct$. In addition to BHs, there is dark matter inside and around the cluster of BHs, provided that BHs do not constitute all the dark matter. The cluster represents the initial density perturbation (of the entropy or isocurvature type) which determines the subsequent evolution of this composite system, and this perturbation competes with ordinary inflationary perturbations which produce CMB anisotropy and most of the large scale structures as in the standard cosmological scenario. The initial mass profile of the BHs cluster $M_h(r_i)$ is shown in Fig.~\ref{massprof1} in comparison with dark matter mass profiles $M_{\rm DM}(r_i)$ inside and around the same cluster. 

Let us consider the spherical shell with the total BHs mass $M_h$ inside the sphere of radius $r$, radiation density $\rho_r$, dark matter density $\rho_{\rm DM}$, and Lambda-term density $\rho_{\Lambda}$. The latter quantity is important only at the nearest epoch $z\sim1$ and can be neglected at the early stages $z\gg1$ of the Universe evolution. The radiation inside the cluster is homogeneous with high accuracy, and its gravitation can be taken into account by the substitution $\rho_r\to\rho_r+3p_r/c^2=4\rho_r/3$ into Newtonian equations, and analogously for the Lambla-term. The evolution equation for the shell has the form
\begin{equation}
\frac{d^2r}{dt^2}=-\frac{G(M_h+M_{\rm DM})}{r^2}-\frac{8\pi G\rho_r
r}{3}+\frac{8\pi G\rho_{\Lambda} r}{3}. 
\label{d2rdt1}
\end{equation}
Following \cite{KolTka94} we use the parametrization $r=\xi a(t)b(t)$, where $\xi$ is the comoving radius, $a(t)$ is the scale factor of the universe, and the function $b(t)$ describes the inhomogeneous evolution. The DM mass inside the shell is $M_{\rm DM}=(4\pi/3)\rho_{\rm DM}(t_0)\xi^3$, and the function $a(t)$ obeys the Friedmann equation $\dot a/a=H_0E(z)$, where the red-shift $z=a^{-1}-1$, $H_0$ is the Hubble parameter at the present time, and the function 
\begin{equation}
E(z)=[\Omega_{r,0}(1+z)^4+\Omega_{m,0}(1+z)^3+
\Omega_{\Lambda,0}]^{1/2}. \label{efun}
\end{equation}
By using the Friedmann equation the (\ref{d2rdt1}) can be rewritten as 
\begin{equation}
\frac{d^2b}{dz^2}+\frac{db}{dz}S(z)+
\left(\frac{1+\delta_h}{b^2}-b\right)\frac{\Omega_{m,0}(1+z)}{2E^2(z)}=0,
\label{d2bdz1}
\end{equation}
where
\begin{equation}
S(z)=\frac{1}{E(z)}\frac{dE(z)}{dz}-\frac{1}{1+z},
\end{equation}
and $\delta_h=M_h/M_{\rm DM}$. At the limit $\rho_{\Lambda}\to0$ the Eq.~\eqref{d2bdz1} is analogous to the equation obtained in the work \cite{KolTka94} for the case of axionic dark matter. But we consider much more higher densities ($\delta_h>10^4$) where the approximate fitting solutions of \cite{KolTka94} are not applicable, so one should solve Eq.~(\ref{d2bdz1}) numerically. We start the numerical solution from the moment of the horizon crossing of each particular shell with the initial conditions shown in Fig.~\ref{massprof1}.

Some particular shell stops expanding, then $\dot r=0$ ($db/dz=b/(1+z)$) at some radius $r=r_s$. After the contraction from $r_s$ to $r_c=r_s/2$ the shell virializes and fixes finally at the radius  $r_c$. Therefore the mean density $\rho$ of the formed cluster is 8 times its density at the moment of maximum expansion
\begin{equation}
\rho=8\rho_{m,0}(1+z_s)^3b_s^{-3},
\end{equation}
and the virial radius is equal to
\begin{equation}
r_c=\left(\frac{3M_{\rm DM}}{4\pi\rho}\right)^{1/3}.
\label{rceq}
\end{equation}

The shells detachments start from the center of the cluster. 
At the very early stage of evolution the central mass $M_c\simeq1.1\times10^4M_{\odot}$ have occurred under the Schwarzschild radius. This is the mass of the central BH which can grow further due to the smaller BHs capture and baryons accretion.

The shells with $\delta_h>1$ ($M_{\rm DM}<M_h$) detached at the radiation dominated stage, but the remaining shells detached later at the matter dominated stage. For the outer shells the dark matter has the principle influence on the gravitational dynamics. 

\begin{figure}[t]
	\centering
	\includegraphics[width=0.7\textwidth]
{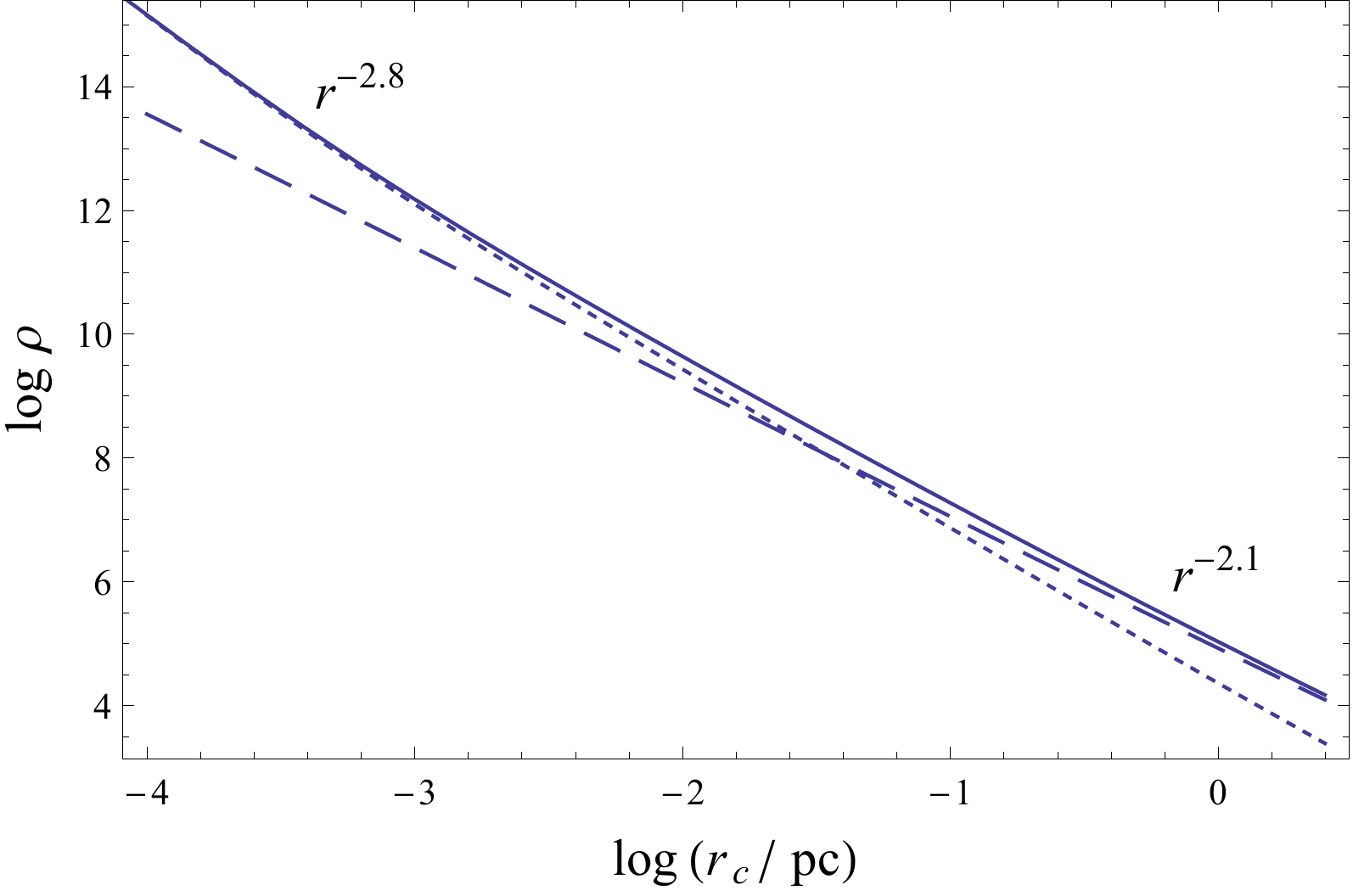}
	\caption{\label{massprof2} Density profile (\ref{dprofeq}) for dark matter (dashed line), BHs (dotted line), and for the total density (solid line). The asymptotic power-law lines are shown. The density and radius from the center are in $M_{\odot}$pc$^{-3}$ and pc unints, respectively.}
\end{figure}

The dark matter accompanies the BHs shells detachments with the formation of the density profile in analogy with secondary accretion mechanism. But the density profile does not follow the usual secondary accretion low $\rho\propto r^{-9/4}$ due to the non-compactness of the seed mass
\begin{equation}
\rho_{\rm DM}(r)=\frac{1}{4\pi r^2}\frac{dM_{\rm DM}(r)}{dr},
\label{dprofeq}
\end{equation}
where $M_{\rm DM}(r_c)$ is obtained from the solution of (\ref{d2bdz1}). This density profile is shown in Fig.~\ref{massprof2}.

The virialized region is including more and more DM shells until some outer shell is distorted by the gravitation from the surrounding density perturbations of inflationary origin. The equality of the counteracting forces from the total inner mass and from the outer perturbations determines the final radius of the system. The accretion of the gas onto the central BH can explain the early activity. And at the present epoch these objects look like dense galaxies with dormant BHs at the centers.